\documentclass[11pt]{article}
\usepackage{epsfig}
\usepackage{geometry}                
\usepackage{graphicx}
\usepackage{amssymb}
\usepackage{epstopdf}
\title{An analysis of spectra in the Red Rectangle nebula}

\author{Fr\'ed\'eric Zagury \thanks{e-mail:fzagury@wanadoo.fr}
\\Institut Louis de Broglie, 23 rue Marsoulan, 75012 Paris, France}

\begin{document}
\maketitle

\begin{abstract}
This paper presents an analysis of a series of spectra in the Red Rectangle nebula.
Only the reddest part of the spectra can 
safely be attributed to light from the nebula, and indicates 
Rayleigh scattering by the gas, in conformity with the large angles of 
scattering involved and the proximity of the star.
In the blue, light from HD44179, refracted or scattered in the 
atmosphere, dominates the spectra.
This paper questions the reliability of ground-based observations of 
extended objects in the blue.
\end{abstract}
    \section{Introduction}
This paper will pursue the investigations initiated in a preceding 
article (Zagury, 2006) 
on the effects of atmospheric extinction on starlight.

Long slit spectroscopy of a sample of objects (stars, nebulae, 
galaxies) was used to 
determine the nature of the light received by a telescope at close 
angular distances from a star, before the spectrum reaches the night sky 
spectrum.
Direct light from the star, refracted by the 
atmosphere, will be detected over the first few arcseconds.
Over the next few tens of arcseconds, forward scattering of starlight by aerosols 
will dominate the spectrum, until it fades out under the night sky 
spectrum.

Light from the central star, refracted or scattered in the atmosphere,
can not be neglected when dealing with spectra 
of a nebula, since it will compete with the light 
scattered by the nebula. 
The continua of the spectra of NGC6309, 
NGC6891, and NGC2022, for instance, as they are given 
in the Kwitter-Henry database
(http://oit.williams.edu/nebulae/browse.cfm),
more likely  reflect the spectrum of the central star,
than the true nebular spectra (Zagury, 2006). 

If this preceding paper emphasized the importance of atmospheric extinction 
for our understanding of ground-based observations of nebulae, 
it was not quantitative enough to
separate, in the observed spectrum of a nebula, atmospheric effects 
from true nebular spectrum.
This problem will guide the present paper, 
which will focus on spectra of the 
Red Rectangle nebula, observed at different positions in the 
nebula, and at different dates.
\section{Observations} \label{data}
The observations of the Red Rectangle used in this article come 
from three observing runs at Fred L. Whipple Observatory,
in December 2001, February 2002, and March 2003.
Spectra were acquired with the FAST spectrograph, mounted on the Tillinghast 1.5 
meter telescope.
Technical informations on FAST will be found in Fabricant \emph{et al.} 
(1998).

FAST is a long slit spectrograph, 
which samples a region 3' long and 3'' wide on the sky and 
covers (in the configuration used for the observations) the [$3660\,\rm 
\AA$, 
$7530\,\rm \AA$] 
wavelength range, with a spatial resolution of 0.61"/pixel.
The slit is oriented E-W.
Different positions of the slit across the 
nebula are thus characterized by their declination offset from HD44179.

The 2-D spectra presented here were binned by two (December 2001, 
March 2003) or by four (February 2002), along their spatial dimension,
during the data reduction 
process (Tokarz \& Roll 1997), yielding a final angular resolution of 
1.2" or 2.4" per pixel.
1-D spectra of stars are extracted from the 2-D arrays by adding the spectra 
far above the background.
The wavelength resolution is $1.4\,\rm\AA$/pix.
The seeing during the three observing runs was good, 1-2".

\begin{figure*}[]
\resizebox{1.\textwidth}{!}{\includegraphics{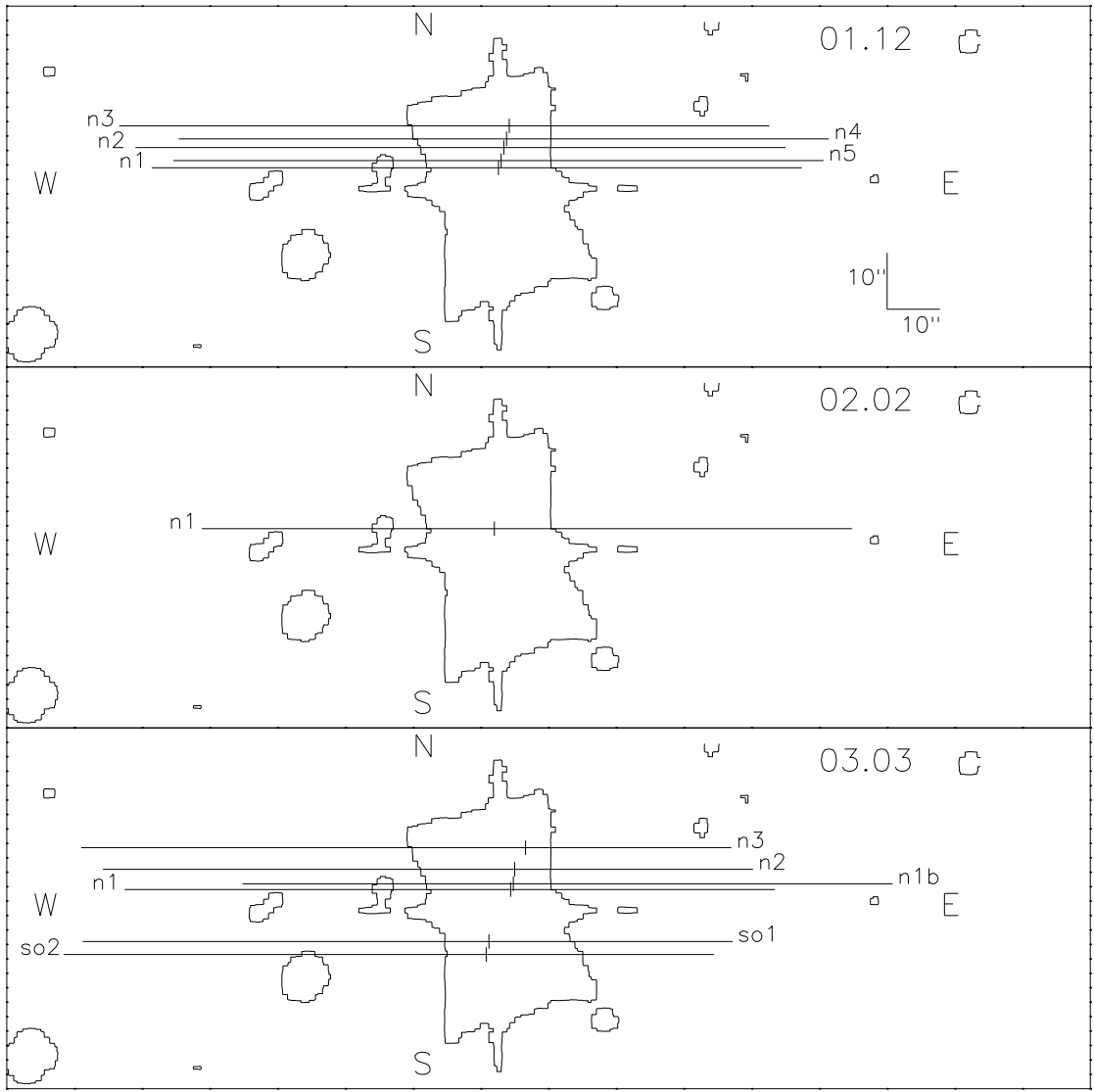}} 
\caption{Summary of the observations of the Red Rectangle nebula.
For each run (one plot per date), the successive positions 
of the slit (horizontal lines) are projected on a contour plot of the DSS2 
red survey image centered on HD44179 
($\alpha=06^{h}¥19^{m}¥37^{s}¥$, $\delta=-10^{\circ}38' 15''$, J2000).
The field is 3.6'$\times$1.8' large.
The slit is oriented E-W, i.e. pixel numbers increase with right 
ascension.
On each line, the approximate position of the pixel (of the 2-D array)
with maximum signal is marked by a vertical line. } 
\label{fig:pos}
\end{figure*}

Main informations on the data are grouped in Table~\ref{tbl:obs}.
Figure~\ref{fig:pos} shows the projected positions of the slit on the 
nebula, for each 
run, and specifies the position on the slit of the pixel 
with maximum signal. 

An observation will be designated by its 
position ($n1$, $n2$\ldots), as it is given in Table~\ref{tbl:obs}, 
and the date of observation.
The spectrum at pixel `$x$', extracted from 
the 2-D array of an observation, will be noted `$sx$'.
`$\sigma$' stands for the ozone cross-section, 
and is expressed in cm$^{2}$¥.
Vertical dotted lines on the plots give the  limits of
the ozone absorption region.
   \section{December 2001 observations} \label{dec01}
   \subsection{HD44179} \label{dec01et}
\begin{figure*}[p]
\resizebox{1.\textwidth}{!}{\includegraphics{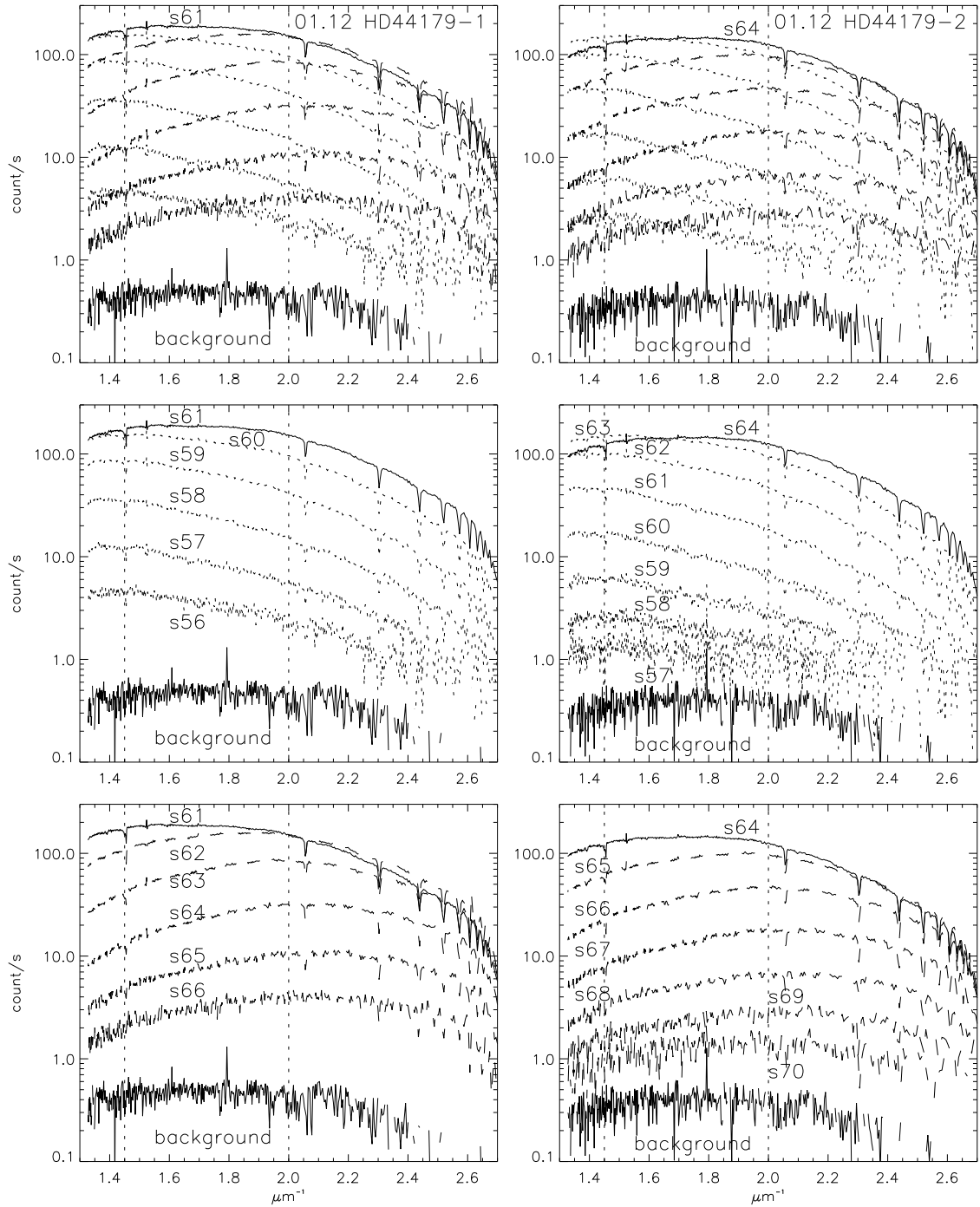}} 
\caption{2-D arrays for the December 2001 first (left column) 
and second (right column) observations of HD44179.} 
\label{fig:dec01et}
\end{figure*}

Spectra extracted from the 2-D arrays of
HD44179's observations, at the beginning and at the end of December 2001 run,
are plotted Figure~\ref{fig:dec01et}.

For each 2-D array a main spectrum is identified ($s61$ for the first 
observation, $s64$ for the second), which fixes the direction 
of HD44179.
On each side of the main pixel the intensity decreases steeply, to 
less than $\sim 1$~count/s, over $\sim 7$~pixels or $\sim 8$''.
The shape of the spectra with level above $1$~count/s evolves in a regular
way, with a systematic increase of the average slope of the 
spectra with increasing pixel number (see Zagury, 2005).
Some spectra, $s65$ and $s64$ in the first observation, $s60$, $s59$, 
in the second one, have a slight bump in the red (the red bump is more 
pronounced in February 2002 observation, Section~\ref{feb02})
which is due to the nebula.
$8$'' away from the direction of HD44179, the spectra have reached a 
shape (mentioned as the background on the plots) which is common to 
the following spectra and to both sides of HD44179.

\begin{figure*}[]
\resizebox{\textwidth}{!}{\includegraphics{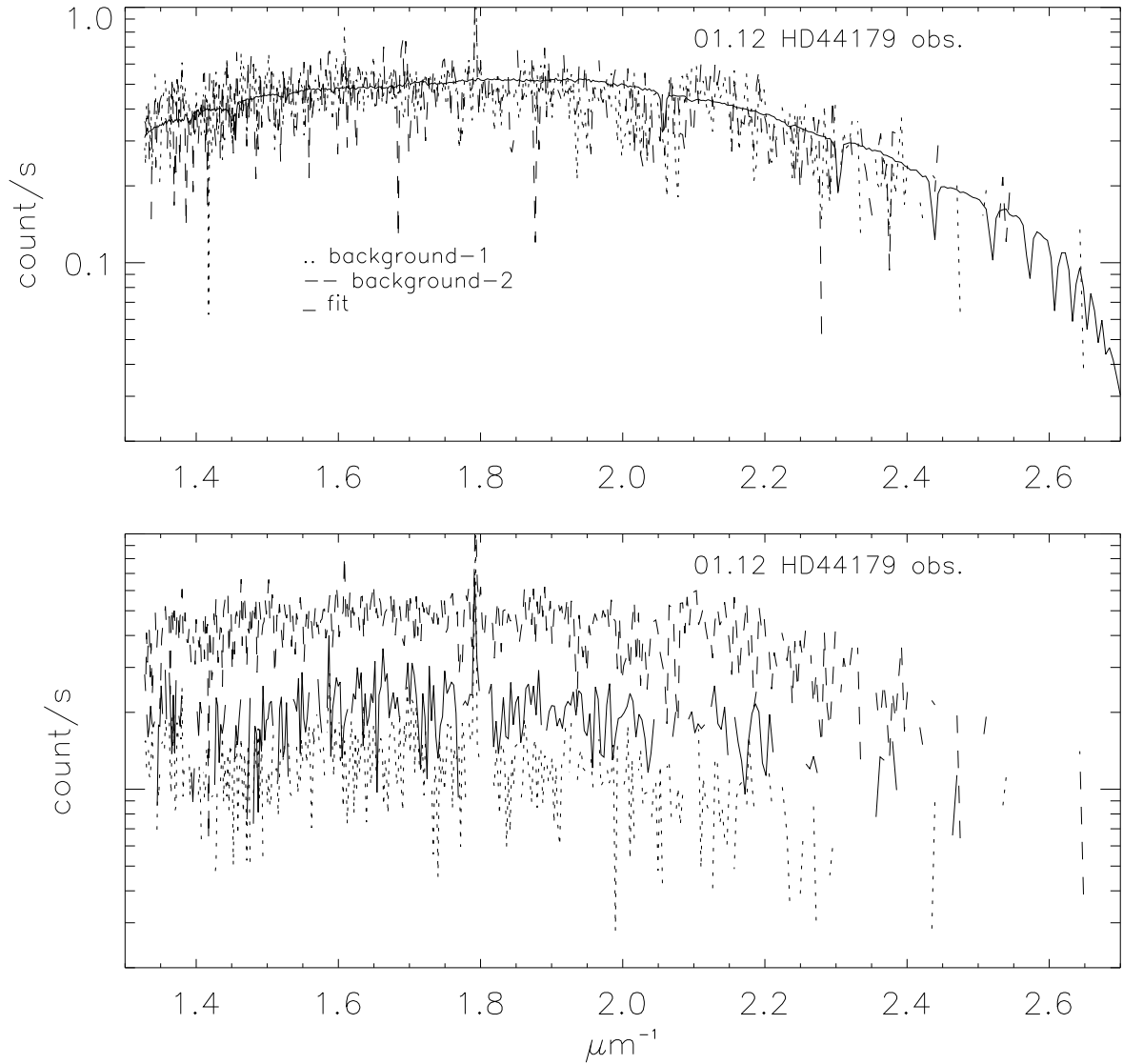}} 
\caption{
\emph{Upper panel}: backgrounds (average of eight spectra of the 2-D arrays, at a mean distance 
of 11" from HD44179) for the two observations of HD44179 have an 
identical shape, proportional to the spectrum of HD44179 times 
$1/\lambda$.
\emph{Lower panel}: decrease of the first observation background with 
distance from HD44179.
The backgrounds are the average of 8, 22, 40 spectra, taken at 
mean distances of $11$", $30$", and $60$", from HD44179.
} 
\label{fig:dec01etfd}
\end{figure*}

The noise level is $\sim 0.2$~count/s, less but not far from the 
level of the background.
The  background spectra of 
Figure~\ref{fig:dec01et} are the average of 8 spectra at a mean 
distance of 11" from HD44179.
The backgrounds, identical in the two observations, 
are interpreted (Zagury, 2006) as light from HD44179 scattered by aerosols in the 
atmosphere, with a spectrum proportional to the spectrum 
of HD44179 and to $1/\lambda$ (Figure~\ref{fig:dec01etfd}).
Moving away from HD44179, the background progressively diminishes in intensity, 
with a constant shape, which will ultimately also be modified, because of the attenuation of 
the scattering and the merging of the night sky spectrum
(Figure~\ref{fig:dec01etfd}, bottom plot).

\begin{figure*}[]
\resizebox{\textwidth}{!}{\includegraphics{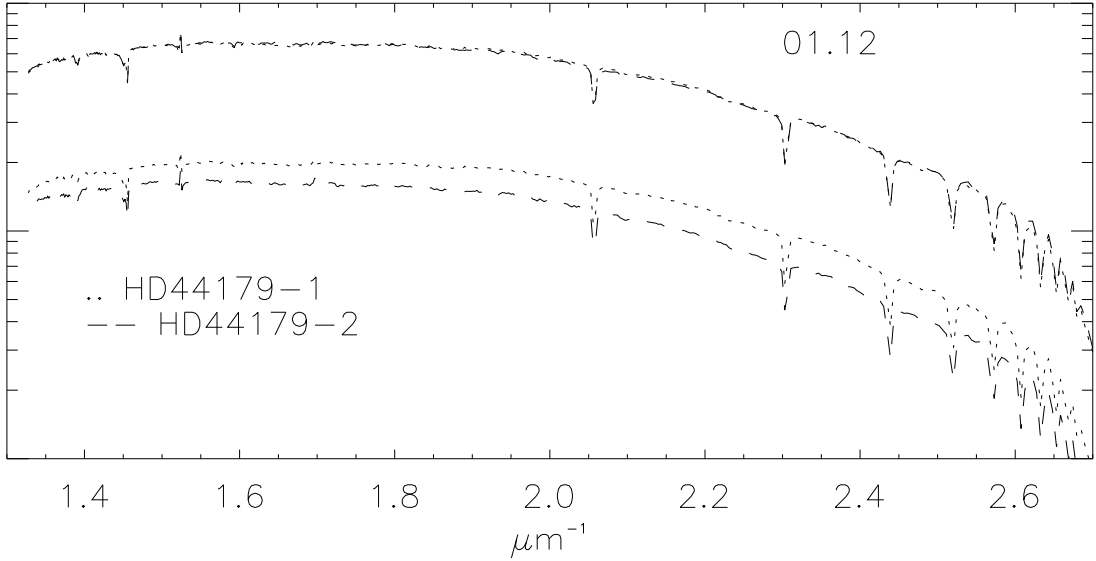}} 
\caption{Comparison of the two December 2001 1-D spectra of HD44179.
\emph{Bottom spectra}: between the beginning (dots) and the end 
(dashes) of the run, the 1-D spectrum of 
HD44179 has decreased in intensity. 
The slope of the spectrum is slightly decreased in the blue.
\emph{Top spectra}: the difference of atmospheric extinction between 
the two observations is corrected by multiplicating the second observation 
1-D spectrum by $1.15e^{0.005/\lambda^4}e^{8\,10^{18}¥\sigma}$.} 
\label{fig:dec01etc}
\end{figure*}

Between the beginning and the end of the run 
the airmass has increased from 1.41 to 1.48.
The result is a decrease of the 1-D spectrum of HD44179 by a factor 
of 1.15, a slightly more pronounced depression in the ozone absorption 
region, and an increase of the bending in the blue 
(a decrease of the blue slope, bottom spectra of Figure~\ref{fig:dec01etc}).
These differences between the spectra can be corrected analytically
(top spectra of Figure~\ref{fig:dec01etc}).
   \subsection{Nebula} \label{dec01neb}
\begin{figure*}[p]
\resizebox{1.\textwidth}{!}{\includegraphics{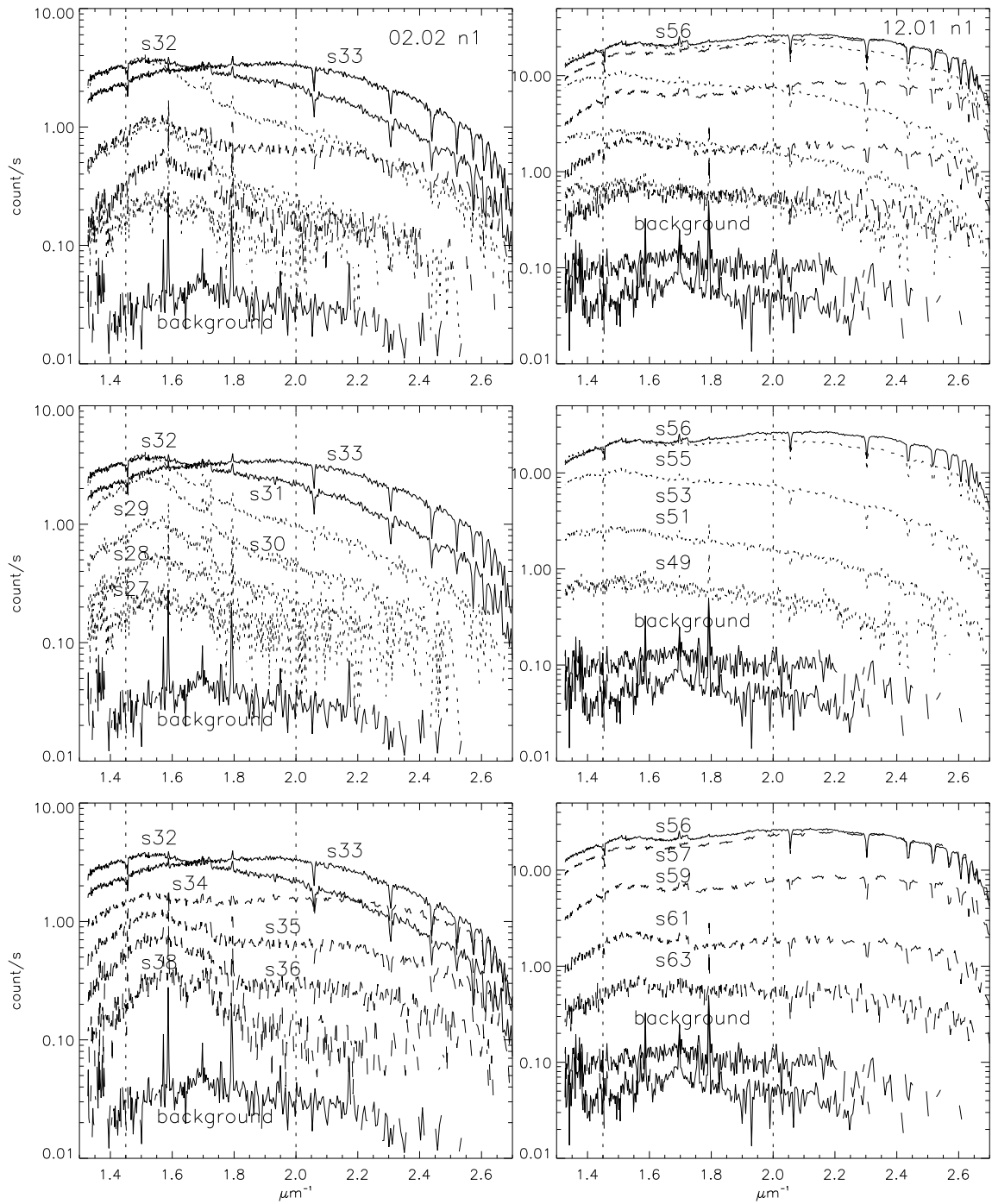}} 
\caption{December 2001 (right) and February 2002 (left) observations of 
position $n1$.
Two backgrounds, close to the nebula and at the edges of the slit,
are plotted for the December 2001 observation, which show the transition
from HD44179's light scattered in the atmosphere to night sky 
spectrum. 
} 
\label{fig:decfebn1}
\end{figure*}

The 2-D arrays of the December 2001 nebular spectra are presented in 
Figures~\ref{fig:decfebn1} (right column) and \ref{fig:dec01n2n3}.
Spectra from positions $n4$ and $n5$, which are similar to $n3$'s 
(see below), were not represented.
The main spectrum for each observation is in plain line.
Its position on the slit, and in the nebula, is marked on 
Figure~\ref{fig:pos}.

The red bump, characteristic of the Red Rectangle, is present in all 
2-D arrays, with a spatial extent that matches the width of the nebula 
at the corresponding declination [derived from the red Digital Sky 
Survey (DSS2) image].

Due to the long exposure times low level background spectra can be 
determined with better accuracy than for the 2-D arrays of HD44179. 
Two background spectra are represented in solid line on each plot, 
one immediately outside the nebula, one close to the extremities of 
the slit.
Backgrounds of all nebular observations, taken 
at the extremities of the slit, are identical 
(in shape and intensity, Figure~\ref{fig:dec01fd}, left).
Their shape matches the night sky spectrum of Massey \& Foltz (2000), 
except in the blue where there is an excess of signal for the 
backgrounds (Figure~\ref{fig:dec01fd}, middle).
This excess is attributed to remaining light from 
HD44179 scattered in the atmosphere, and is well fitted by the 1-D 
spectrum of HD44179 (first observation of the run) times 
$1/\lambda$ (and a constant factor, middle plot of Figure~\ref{fig:dec01fd}).
Subtraction of the scattered light component will 
reproduce the Kitt Peak night sky spectrum 
(right plot of Figure~\ref{fig:dec01fd}).
Backgrounds taken close to the nebula are equal to the 
far-away backgrounds with an additional component of light from 
HD44179 scattered in the atmosphere (Figure~\ref{fig:dec01fd1}).

The first position observed in the nebula, 
$\sim 5$'' north from HD44179, is
close enough to HD44179 to consider a possible contamination of the 
spectra by direct light from HD44179
[as for NGC6309 and NGC6891 in Zagury (2005)].
In the blue, the main spectrum ($s56$, right plot of
Figure~\ref{fig:decfebn1}) of the 2-D array recalls, by its shape 
as well as by its intensity, spectra of the 2-D arrays of HD44179 
taken at approximately equal distances from the star, $s64$ and $s65$ of HD44179 first 
observation, $s66$ and $s67$ of the second observation 
(Figure~\ref{fig:dec01n1et}).
It is (in the blue) exactly proportional to $s65$ of the first observation, and nearly 
proportional to $s67$ of the second one (a small correction for gas 
extinction is necessary to superimpose the spectra in the blue);
the spectrum of the nebula is superimposed, in the red, on the spectrum 
of HD44179.

Ouside the nebula, still at position $n1$, 
the shape of the background spectrum evolves from 
a predominance of light from HD44179 scattered in the atmosphere 
(top background spectra of Figure~\ref{fig:decfebn1}, right column 
plots, with a 
shape identical to the background found for the observation of HD44179) to 
a predominance of the night sky spectrum (bottom background spectra), 
at the edges of the slit.

The other observations of the nebula are farther away from HD44179; 
the proportion of direct light from HD44179 should be highly 
diminished, which explains the important fall
in intensity of the spectra 
(Figures~\ref{fig:dec01n2n3}, \ref{fig:dec01n1et}, and \ref{fig:dec01nc}).

Main spectra at positions 3, 4 and 5 have rigorously identical 
shapes; they differ by a constant factor close to 1 
(dotted spectra of Figure~\ref{fig:dec01nc}).
The blue part of $n3$, $n4$, $n5$, 2-D arrays 
keeps a constant shape in and out of the nebula (Figure~\ref{fig:dec01n3bleu}),
with only a diminution of the magnitude when moving away from it.
The spectra are also observed at about the same distance (8"-15") from HD44179 where 
the backgrounds for the HD44179 observations were established (Section~\ref{dec01et}).
In the blue they are similar to these backgrounds, both in shape 
and in magnitude (a few 0.1~count/s).
The backgrounds from the observations of HD44179 would 
superimpose well, with a scaling factor close to 1, on the blue part 
of the spectra of Figure~\ref{fig:dec01n3bleu}.
On the figure, the backgrounds are 
fitted, in the blue, by the spectrum of scattered light (by atmospheric aerosols) from 
HD44179 (proportional to the product of the spectrum 
of HD44179 by $1/\lambda$).

Spectra from $n3$, $n4$, $n5$, 2-D arrays, in the nebula, 
are thus made of two components: 
nebular light in the red, 
stacked on scattered light (from HD44179) in the atmosphere, which 
dominates in the blue.
The red rise is constant for the closest positions to the main pixel, 
but its slope will 
diminish towards the edges of the nebula, because of the increasing 
proportion, in the spectra, of scattered light in the atmosphere.

The main spectrum ($s52$) of the 2-D array at position $n2$ is of higher level. 
The red rise is as for the main spectra at 
positions 3, 4, and 5, but the spectra differ in the blue
(Figure~\ref{fig:dec01nc}).
The blue part of $s52$ is in-between, and proportional to
spectra $s67$ and $s68$, from the first observation of HD44179 
(Figure~\ref{fig:dec01n2b}), observed at about the same angular distance from 
HD44179  as $s52$ (of $n2$) is.
The blue parts of the other spectra of $n2$ 2-D array are either like  
that of $s52$ (for pixel numbers close to 52), or, 
like the blue part found previously at positions $n3$, $n4$, $n5$.

It will be concluded that for all spectra in the nebula, only the red 
part of the spectra can safely be attributed to light from the nebula.
In the blue, the spectra are dominated either by direct, refracted, light from 
HD44179, or by light from HD44179 forward scattered by atmospheric aerosols.

One can wonder why $s52$ of $n2$ is three 
times larger than $s48$ of $n5$ (Figure~\ref{fig:dec01nc}), while position $n5$ is closer to HD44179 than 
$n2$ is.
The reason might be the increase of airmass between the two 
observations (Table~\ref{tbl:obs}), which diminishes the amount of direct light from 
HD44179 and increases the scattered light in $n5$ observation.
   \section{February 2002 observations} \label{feb02}
There are three February 2002 observations, one of HD44179, one 
$5.4$'' north of HD44179 (position~1), and one of HD44113.

The two star observations are plotted on Figure~\ref{fig:feb02et}.
The decrease of the signal, when moving away from the 
main pixel ($s35$ and $s36$ respectively for HD44179 and HD44113, 
solid line on the plots),
is similar in both observations, although it is faster for HD44179 
(while HD44179 and HD44113 have comparable magnitudes): 
it occurs within 1-2 pixels ($\sim 4$") from $s35$ for HD44179, 
within 4 pixels ($\sim 8$") for HD44113 (as for the December 
2001 observation of HD44179).
This may be related (Zagury, 2006) to the exposure 
time, which, for HD44179, has been reduced from 15 seconds in December 
2001 to 5 seconds in February 2002.
The exposure time for HD44113 is 15 seconds, as it was for the 
December 2001 observations of HD44179.

The spectrum of the nebula clearly appears in the red, superposed 
on the spectrum of HD44179, for spectra $s33$ and $s32$, $s37$ and 
$s38$.
The nebular contribution to the spectra is far more significant in 
this observation than it was for December 2001 ones, which, a priori, can be 
due either to a slightly different position of the slit axis ($3$'' wide) 
in declination, or, more likely,
to a difference in atmospheric extinction between the two 
nights [which implies a lower extinction for February 2002 
observations and could be related either to the lower airmass
(Table~\ref{tbl:obs}), or, to the smaller exposure time of these 
observations]. 
The nebular spectrum is present in the red only 
and does not seem to perturbate the spectrum of HD44179 in the blue.

The nebular 2-D array was presented, along with the one of December 
2001 observation at the same declination, in Figure~\ref{fig:decfebn1}.
Two main spectra ($s32$ and $s33$) merge out from the 2-D array 
(rather than one in December 2001 observations), 
which is to be attributed to the binning by four 
procedure used in the data reduction of February observations.
The shape of $s33$ (a fit will be proposed Section~\ref{comp}), 
and the decrease of the spectra on each side of the 
central pixels, indicate, as for $n1$ December 2001 observation, that the 
spectra are dominated in the blue by direct light from HD44179:
the nebular spectra are superposed on HD44179's spectra, and restricted 
to the red.
The background reaches the night sky spectrum at the edges of the slit.
   \section{March 2003 observations} \label{ma03}
   \subsection{HD44179} \label{etma03}
The two March 2003 observations of HD44179 are similar, and were discussed 
in Zagury (2005) (Figures~1 and 2).
As for February 2002 observation, direct light from HD44179 is 
spread over $\sim 8$", i.e. $\sim4$" (3-4 pixels) on each side of the central pixel.
The noise level is $\sim 1$~count/s, higher than in December 2001, 
which is due to the smaller exposure time.
   \subsection{Nebula} \label{neb03}
On Figure~\ref{fig:ma03neb}, March 2003 observations of the Red Rectangle nebula are
classified, from left to right and top to bottom, 
by decreasing importance of the signal. 
Main features are the same as for December 2001 observations.

The backgrounds away from the nebula are identical 
(Figure~\ref{fig:ma03fd}, left), in shape and intensity.
The match with Massey \& Foltz (2000) night sky spectrum is good except in the blue. 
The excess found in the blue
reproduces the background of HD44179 observations;
it is proportional to the product of HD44179 1-D spectrum (first 
observation) by $1/\lambda$, and is interpreted 
as remaining light from HD44179 scattered in the atmosphere
(Figure~\ref{fig:ma03fd}, right).

Main spectra at positions $n2$, $n3$, and $so2$ are identical, 
up to a constant factor (Figure~\ref{fig:ma03n2n3}).

In the red (Figure~\ref{fig:ma03neb}), the red bump of the Red Rectangle 
is present, within the nebula, in all observations.  
The red rise keeps a constant slope around the main pixel of an 
observation, and will diminish toward the edges of the nebula; the bump 
disappears when the pixel is out of the nebula.

The red rises of the main spectra of all observations (except $n1$) are 
proportional (Figure~\ref{fig:ma03r}, left).
$n1$ main spectrum shares the same slope in the reddest part of the 
red rise, 
but is less steep from the ozone absorption region onwards 
(Figure~\ref{fig:ma03r}, right).

The shape of the blue spectra, for a given 2-D array of one of the four observations 
with lower signal 
($n2$, $n3$, $so1$, $so2$), is the same in and out of the nebula,
down to the background (Figure~\ref{fig:ma03neb}).
All four observations blue spectra are proportional (Figure~\ref{fig:ma03n2o2}).
One will deduce that, as for December 2001 observations, 
the blue part of these spectra is mainly light from 
HD44179 scattered in the atmosphere (Figure~\ref{fig:ma03n2o2});
and that the difference between the spectra at $so1$ and at positions $n2$,
$n3$, $so2$, is due to the relative proportion in the spectra of  light from 
HD44179 scattered in the atmosphere.

The blue decrease of the main spectrum ($s69$) at position $n1b$ perfectly 
reproduces the blue part of spectrum $s54$ (scaled by a factor of 0.1) 
of the first observation of HD44179 (Figure~\ref{fig:ma03o3}).

$n1$ (Figure~\ref{fig:ma03neb}) is the nebular 
observation with the highest signal 
and the closest to HD44179.
The nebula bump is perceived from pixel 43 to pixel 56.
The main spectrum is $s48$, which is flat in the blue.
Such a flat blue slope is observed in the February 2002 observation 
of HD44179 ($s38$, Figure~\ref{fig:feb02et}, for instance), 
and can be attributed to direct light from HD44179.
Spectra close to $s48$ differ by a gas extinction factor only 
($\propto e^{-g/\lambda^{4}¥}$, Figure~\ref{fig:ma03n1}), as it was shown 
to be the case with the spectra from HD44179's 2-D array (Zagury, 2006).
Away from pixel 48, the spectra match that of other 
observations in the nebula, observed at comparable distances from 
HD44179:
spectrum $s52$ of $n1$ for instance is identical (with a scaling factor of 1.3) to 
the main spectrum, $s69$, of $n1b$,
the blue part of which was shown to be proportional to $s54$ from the first 
observation of HD44179 (Figure~\ref{fig:ma03o3}).

As for December 2001 or February 2002, 
the spectra of the nebula observed in March 2003
reveal the light from the nebula in their reddest part (where atmospheric 
extinction is minimized, before ozone absorption), but are, in the 
blue (after the ozone absorption region), dominated by light from 
HD44179 -direct light or light scattered in the earth atmosphere. 
   \section{Comparison of the observations at different dates} \label{comp}
Figure~\ref{fig:fdc} compares the backgrounds, observed away from 
HD44179, of the three observing runs.
In the red, a scaling factor close to 1 will give a good superimposition 
of the spectra.
In the blue, the three backgrounds differ
by the amount of light from HD44179 scattered in the atmosphere.
The difference between December 2001 and March 2003 
backgrounds is small.
It becomes significant with the February 2002 background, which is 
understandable since only the closest position to HD44179 was observed 
in February 2002.
We conclude that the night sky was nearly the same for the three runs.
It implies that the conditions of observation, the sensitivity of 
the detector and the atmospheric extinction are comparable.

The 1-D spectra of HD44179 observed at different 
dates superimpose well after correction for a 
difference of Rayleigh extinction (Figure~\ref{fig:etc}) between the observations.

Main spectra at positions $n2$ and $n3$ 
are proportional in March 2003 (Section~\ref{ma03}), while they have 
different shapes in December 2001 observations (Section~\ref{dec01}).
The pointing of the telescope could be responsible for part of this 
discrepancy, but the most reasonable explanation is that the 
difference of optical depths (airmass) between the two dates changes 
the proportion of refracted and scattered light, from HD44179, in the 
spectra. 
It means that the standard way of reducing data, subtraction of a 
background and normalization by the spectrum of a reference star, can 
not be applied to these observations.

A precise position-to-position comparison of December 2001 and March 2003 
observations (Figure~\ref{fig:nloin}), for positions $n2$ and $n3$, 
shows that the red rise of the nebular spectra is identical from one 
date to the other,
while the spectra differ in the blue from the ozone 
absorption region onwards.

$n1$ is the nearest position to HD44179 observed in the nebula, and 
the one for which direct light from HD44179, refracted in the 
atmosphere, enhances the level of the spectrum far above that at the 
other positions.
It is also the position for which differences between the different 
dates are the most striking (Figure~\ref{fig:n1}).
The exact position of the slit axis with respect to HD44179 
can influence the level of the spectrum.
The high airmass (A.M=1.51, against 1.41 for December 2001 
and 1.38 for February 2002) probably contributes to the 
low level of March 2003 spectrum, and its steep rise in the red, 
which is close to what is observed at larger distances from HD44179:
an increase of extinction 
diminishes the direct light from HD44179, which is the main component of the 
$n1$ spectra in the blue, and the reason for its high level;
the Red Rectangle nebular bump is then highlighted.

This further raises the question of the influence of the blue 
continuum on the slope of the red rise at $n1$, 
for December 2001 and February 2002 observations.
Figure~\ref{fig:n1} (right plot), for instance, shows  that it is possible to reproduce 
February 2002 $n1$ spectrum by the sum of spectra extracted from the 2-D 
arrays of December 2001 observations of HD44179 and of position $n2$.
We also note, still concerning the 2-D array of February 2002 observation,
that when the blue continuum diminishes, the slope of 
the red rise increases to its value at positions 
farther away from HD44179 ($s28$ of $n1$, February 2002 
observation, for instance, is identical to the main spectrum $s49$ of 
$n2$, December 2001 run). 
   \section{Discussion} \label{dis}
The previous analysis leads to two first conclusions.
One is that the only part of the spectra which can be attributed to 
light from the nebula is the red bump, the spectrum at smaller 
wavelengths being light from HD44179 refracted or scattered in the 
atmosphere.
The second conclusion is that only the red rise of the bump is not modified from 
one observation to another.

What is the meaning of the bump, and where is 
the light from the central star scattered by the nebula? 
Either the light from HD44179 scattered by 
the nebula really is reduced to a small offset, 
in the blue and maybe in the red; 
or, it is seen in 
the red only because it is strongly extinguished from the ozone absorption 
region onwards (and towards the UV).   

In the first case, the red bump could be, as it has been admitted 
so far, an emission process due to 
some kind of particle which would absorb light in the UV and 
fluoresce in the red.
Light from HD44179 scattered by 
the nebula is then negligible compared to the fluorescence process.
However, no such particles have ever been synthetized on earth with 
the required spectrum, and 
their nature is still today, after some thirty years of research, 
a speculative subject (Li~\& Draine 2002; Van Winckel, Cohen \& Gull 2002).

Also in this case, 
the red bump could result from extinction and scattering by large grains or by the 
gas [such an explanation was suggested  by Cohen \emph{et al.} (1975), see 
Section~\ref{rr} (Appendix)].
Forward scattering by large grains would give a spectrum $\propto 
e^{-a/\lambda}/\lambda$ ($a$ a constant), with a much too large bump to fit the data 
[all spectra of the Red Rectangle nebula found in the litterature 
give a ratio of FWHM to central wavelength in the range 0.18-0.30, 
while scattering by large grains would imply a ratio close to 4].
It is also not compatible with the large angles of scattering 
involved in the present case.
Scattering by gas will have a spectrum 
$\propto e^{-g/\lambda^{4}¥}/\lambda^{4}¥$ (Rayleigh, 1871),
which implies a constant ratio of FWHM to 
central wavelength of $0.66$, also too large to match the observations. 

In addition to the 
absence of scattered light from the nebula in the blue, 
if the red bump is a true nebular feature 
(an emission process or due to scattering), it is still 
necessary to understand the reasons for the variability of the blue 
decrease of the bump (Figure~\ref{fig:dec01nc}, Figure~\ref{fig:ma03r}, 
Figure~\ref{fig:ma03n1}, Figure~\ref{fig:nloin}), while the red rise keeps, in general, a constant 
slope (modifications of the red slope are observed close to HD44179 
only, 
and may be due, as pointed out in Section~\ref{comp}, to the underlying continuum of HD44179 light, refracted or 
scattered in the atmosphere).
The correlation between
the modifications of the blue side of the bump, and the ozone absorption 
region, would have to be pure coincidence, as would have to be
the transformations used to 
superimpose the main spectra of $n1$ March 2003 observation 
(Figure~\ref{fig:ma03n1}), which attest the role of atmospheric 
extinction on the decrease.

On the other hand, atmospheric extinction would well explain the 
particularities of the Red Rectangle nebula spectra.
The red rise of the bump is constant, within the nebula, in all 
observations.
It corresponds to a wavelength region with no ozone absorption 
and where Rayleigh extinction is low.
The progressive interruption of the red rise, and the blue decrease, 
both coincide with the 
wavelength region where ozone absorption operates.
At larger wavelengths, Rayleigh extinction by nitrogen ($\propto 
e^{-g/\lambda^{4}}$)¥ will relay 
ozone absorption. 

If atmospheric extinction is the explanation of the red bump, 
the true meaning would be that atmospheric extinction does not act in 
the same way for the nebular light as it does for point sources' light, 
for HD44179 or HD44113 for instance. 
More generally it implies that extended objects are not extinguished by 
the atmosphere in the same way as point sources are,
atmospheric extinction being far more efficient on the former.
This can only be related to turbulence, and to the structure of the atmosphere,
already responsible for such phenomena as scintillation, 
which affects starlight and not the light from planets.

This hypothesis is further supported by the observations presented in 
Zagury (2005).
Faint and extended galaxies observed with FAST 
(see the FAST database at: http://tdc-www.harvard.edu/archive/), as UGC11917,
also show a strong decrease from the ozone absorption 
region onwards, and their blue shape is close to that of the night sky.
Except for the three nebulae used in Zagury (2005), I didn't have access 
to the 2-D spectra of the Kwitter-Henry database, but 
anyone who considers the continua of the Kwitter-Henry spectra 
will observe that they can be classified, according to their shape,
into one of very few categories (I found 4 well defined categories, 
and this classification can probably be simplified).
Many of these spectra have the same marked feature in the red as the Red 
Rectangle has, while others have a star-like spectrum. 
One can surmise 
that this classification results from the relative 
proportion, in the spectra, of light  from the central star refracted 
by the atmosphere, scattered light from the central star 
in the atmosphere, night sky, and nebular light in the red.

The only part of the Red Rectangle nebula
spectrum which is reliable, because of the small 
effect of atmospheric extinction in this wavelength region, is the red 
part, under $\sim 1.5\,\mu\rm m^{-1}¥$ (above $\sim 6600\,\rm \AA$), 
before ozone absorption modifies the spectrum.
To investigate the nature of this part of the spectrum, we need to 
study its power law dependence when
divided by the spectrum of an A0 star (same spectral 
type as HD44179, T$\sim 7500^{\circ}$K¥), as HD44113.
Figure~\ref{fig:pente} shows that the red rise of the nebula spectrum has 
a $1/\lambda^{4}$ dependence, which indicates Rayleigh scattering by the gas.
This would be in conformity with the conditions which prevail for 
these observations, i.e. a scattering medium close to the source of 
light, and large angles of scattering.

The conclusions finally adopted here [that atmospheric extinction is more 
effective on extended objects than it is on point sources, and that the true 
spectrum of the Red Rectangle nebula reflects scattering by the gas 
(hydrogen)] can easily be checked by observing the 
Red Rectangle nebula, and a non-reddened star of A0 type, as HD44113, from 
space, on a large wavelength range.
In the absence of atmospheric extinction, we expect the ratio of the 
spectra to vary as $1/\lambda^{4}e^{-g/\lambda^{4}}$ over all the visible and UV 
wavelength range.  
It would also be desirable to observe different positions in the 
nebula, close to, and far from, HD44179, to determine the variations of the 
intensity of the scattering, and to see if, close to the star, 
scattering by large grains, with a $1/\lambda$ dependence, is 
detected.
   \section{Conclusion} \label{con}
The observations of the Red Rectangle nebula presented in this paper 
outline the importance of atmospheric extinction on the 
re-distribution of starlight in the vicinity ($\sim 1$') of a star.
The spectra of the nebula all appeared to be the superimposition of two 
components, one in the red, clearly associated with the nebula, one in 
the blue, which is mainly either direct light from HD44179 refracted by the 
atmosphere, or light from HD44179 forward scattered by atmospheric 
aerosols (the proportion of direct and scattered lights depends on the 
distance from HD44179, and on the degree of atmospheric extinction). 

In none of the spectra, observed at different places of the nebula, 
and at different dates, was it possible to detect the 
spectrum of the nebula in the blue, which means that either there is 
no (or a negligible) nebular continuum in the blue, or, that atmospheric 
extinction strongly extinguishes the nebular light from the ozone 
absorption region on, towards the UV.

The correlation between the decrease (the blue side of the red bump) 
of the light from the nebula and the ozone absorption region, 
and the variations of the 
spectra at one position from one date to another, naturally favor the 
second hypothesis.
It will then be inferred that faint extended objects are 
more sensitive to atmospheric extinction than point sources are.
Hints to the same conclusion can be found in the spectra of 
the Kwitter-Henry database, 
and of faint extended galaxies observed with FAST.

This is the first and main conclusion of this paper.
It implies that it is not possible to obtain a reliable 
spectrum of faint extended objects, in the visible and in ground-based observations,
by the usual way of normalizing the spectrum 
by the spectrum of a reference star.

These considerations give an alternative solution to the Red 
Rectangle problem, and more generally to the red bump found in the 
spectrum of some nebulae.
To explain the bump, last thirty years of research have focused 
on the existence of particles 
which would absorb UV light and fluoresce in the red (Section~\ref{rr}).
Such particles, which should be abundant enough to account for some $30\%$ 
of the scattered red light in the galaxy (Szomoru \& 
Guhathakurta 1998), with an 
intrinsic photon conversion efficiency of the photoluminescence near 
100$\%$ (Zubko, Smith \& Witt 1999; Van Winckel, Cohen \& Gull 2002), 
would indeed represent a remarkable discovery, 
and it is a puzzling question to understand how particles we are not 
able to synthetize on earth, can be so abundant in the cold and 
nearly empty interstellar medium.
However, none of the attempts to identify these particles has given a satisfactory 
result yet, 
their existence being as speculative today as it was thirty years ago 
(Van Winckel, Cohen \& Gull 2002).

That atmospheric extinction be responsible for the red bump 
would certainly lead to a review of the problem on a more 
simple and pragmatic basis.
It will also explain the similarity outlined by Zagury \& Fujii (2003)
between the spectrum 
of a red horizon at sunrise and the spectra of red nebulae. 
Both spectra result from a combination of Rayleigh scattering by a gas (nitrogen for 
the red horizon, hydrogen for the nebula), and of atmospheric 
extinction (Rayleigh extinction and ozone absorption).
This is the second conclusion of the paper.

One way to verify these two conclusions would be to obtain 
broad-band spectra of extended objects from space, with the Hubble Space 
Telescope for instance.
Concerning the Red Rectangle, I expect the ratio of spectra of 
the nebula and of a non-reddened A0 star, as HD44113, to give a 
$1/\lambda^{4}$ dependence over the complete visible spectrum.
To observe Rayleigh scattering by the gas, observed positions in the nebula 
should be taken far enough from HD44179.
Close to HD44179 it is possible that forward scattering by large 
grains will be more efficient than scattering by the gas, and that a 
$1/\lambda$ dependence be found.
These observations will evidence, for the first time, Rayleigh scattering by the gas 
in the interstellar medium.

A refined study of the spectra from the Kwiter-Henry database, at the early 
stage of data reduction where atmospheric extinction is not yet corrected,
could also provide 
valuable information on the relationship between the conditions of 
observation and the shape of the spectra.
As already mentioned in Zagury (2005), there is little doubt that the 
proximity of the central star, the balance between the proportions of 
light from the 
central star refracted or scattered in the atmosphere, and from the night sky, 
have a determining influence on the final shape of the spectra in the blue.
For nebulae with a reliable spectrum in the red 
(where atmospheric extinction is low), the ratio of 
scattered light from the nebula to the spectrum of an unreddened star of same spectral 
type as the central star, should also provide informations on the 
type of particles which scatter starlight (large grains with a 
$1/\lambda$ dependence, hydrogen for Rayleigh scattering by the gas), 
and thus on 
the geometrical relationship (proximity, angles of scattering) between 
the nebula and the star.

A difference in the effect of atmospheric extinction on point sources 
and on extended objects can only be 
due to the time-dependent structure turbulence introduces in the atmosphere.
The consequences for our understanding of the atmosphere, like the 
relations which exist between the spectra at very short angular 
distances from a star, would merit further attention and investigations.
\appendix
\section{The Red Rectangle} \label{rr}
The large interest (Simbad quotes some 320 references to the Red 
Rectangle) raised in the thirty past years by the Red 
Rectangle takes its origin in the Cohen \emph{et al.} (1975) article, `The peculiar 
object HD44179 ("The Red Rectangle")'.
In conclusion of their paper, the authors address several problems which concern 
the star system at the center of the nebula, the nature of the 
bi-conical nebula, and the reason(s) for its important brightness in the 
red.

Large advances in the comprehension of the inner part of the nebula 
have been made in the 1990's, from high resolution 
imaging, with Roddier \emph{et al.} (1995) who showed that 
we do not see direct starlight from HD44179, only
scattered light
escaping from the poles of an optically thick disc, seen edge-on.
The disc, more likely a torus (Tuthill \emph{et al.} 2002), surrounds a 
spectroscopic binary with an orbital period of 300 days (Van Winckel, Walkens \& Waters 1995).
The optical light escaping the torus is that of an A0 star, with 
effective temperature $\sim 7500^{\circ}$K  (Waelkens \emph{et al.} 1996).¥ 
A schematical representation of this inner system ($\sim 100$~mas 
around HD44179) can be seen in Figure~3 of Waelkens \emph{et al.} (1996) or 
Figure~8 of Men'shchikov \emph{et al.} (1998).
Recent high resolution images (Tuthill \emph{et al.} 2002; Cohen \emph{et al.} 2004) of this inner 
part of the nebula seem to confirm this representation.

On a larger scale the nebula is biconical, and occupies $\sim 
40"\times 40"$ on the sky.
The question of its red color,  
addressed by Cohen \emph{et al.} (1975), is still a subject of debate.
The quite logical proposal by Cohen \emph{et al.} (1975), that it could be light 
from the central A0 star extinguished (by small and large grains) and scattered 
by the gas (because of the large angles of scattering), 
did not attract much attention (except for 
two papers, Greenstein~\& Oke 1977; Perkins \emph{et al.} 1981), on-going research 
being oriented towards the search of particles which would absorb 
in the UV and luminesce in the red (Duley 1985; d'Hendecourt \emph{et al.} 
1986; Seahra \& Duley 1999).
However, the existence of these particles is still in the speculative 
domain (Li~\& Draine 2002; Van Winckel, Cohen \& Gull 2002).
\section*{Acknowledgments}
This work was made possible thanks to the courtesy of Roc Cutri, and John 
Huchra. It is most of all indebted to Lucas Macri, for his unfailing 
kindness in repeating the observations of the Red Rectangle.

{}
\clearpage
\begin{table*}[h]
\caption[]{Observations}		
       \[
    \begin{tabular}{cccccc}
\hline
Position &  U.T.$^{(1)}$& $\Delta t^{(2)}$  & 
$\Delta \delta^{(3)}$ & A.M.$^{(4)}$ & Alt.$^{(5)}$\\ 
\hline
\multicolumn{6}{c}{December 2001 (2001-12-22)}\\
HD44179-1 &08:44:21 &  15 &  & 1.41 &  
44.92 \\ 
n1 &08:45:08 & 60 & 5.4 & 1.41 & 44.80  \\ 
n2 & 08:46:53 &   120 &   9.8&   1.42 &   44.57 
 \\ 
n3 & 08:57:00 &   240 &   14.3 &   1.45 &   43.42  
 \\ 
n4 & 09:03:31 &  120 &  11.5 &  1.46 &  42.96  
 \\ 
n5 & 09:07:44 &  120 &  7&  1.48 &  42.52  
 \\ 
HD44179-2 & 09:10:04 &  15 &   &  1.48 &  
42.44 \\ 
\hline
\multicolumn{6}{c}{February 2002 (2002-02-09)}\\
HD44113 & 5:05:06 &  15 &    &  1.38 &  46.28 \\ 
HD44179 & 5:07:24 &  5 &    &  1.37 &  46.59 \\ 
n1 & 5:14:08 &  60 &  5.3 &  1.38 &  46.15 \\ 
\hline
\multicolumn{6}{c}{March 2003 (2003-03-26)}\\
HD44179-1 & 3:10:46 &  5 &   &  1.51 &  41.40 \\ 
n1 & 3:11:46 &  60 &  5.4 &  1.51 &  
41.18 \\ 
n2 & 3:13:23 &  120 &  9.8 &  1.52 &  
40.86 \\ 
n3 & 3:15:59 &  240 &  14.3 &  1.54 &  
40.30 \\ 
so1 & 3:20:33 &  120 & -5.3&  1.55 &  
39.97 \\ 
n1b & 3:23:11 &  240 &  6.7&  1.57 &  
39.36 \\ 
so2 & 3:28:09 &  240 & -8.3 &  1.59 &  
38.71 \\ 
HD44179-2 & 3:33:27 &  5 &   &  1.60 &  38.52 \\ 
\hline
\end{tabular} 
 \]
\begin{list}{}{}
\item[$(1)$] UT HH:MM:SS at start of exposure.
\item[$(2)$] Duration of exposure (seconds).
\item[$(3)$] Elevation offset from HD44179.
\item[$(4)$] Air mass.
\item[$(5)$] Altitude ($^{\circ}$)¥ of the object at time of observation.
\end{list}
\label{tbl:obs}
\end{table*}
\clearpage
\begin{figure*}[p]
\resizebox{1.\textwidth}{!}{\includegraphics{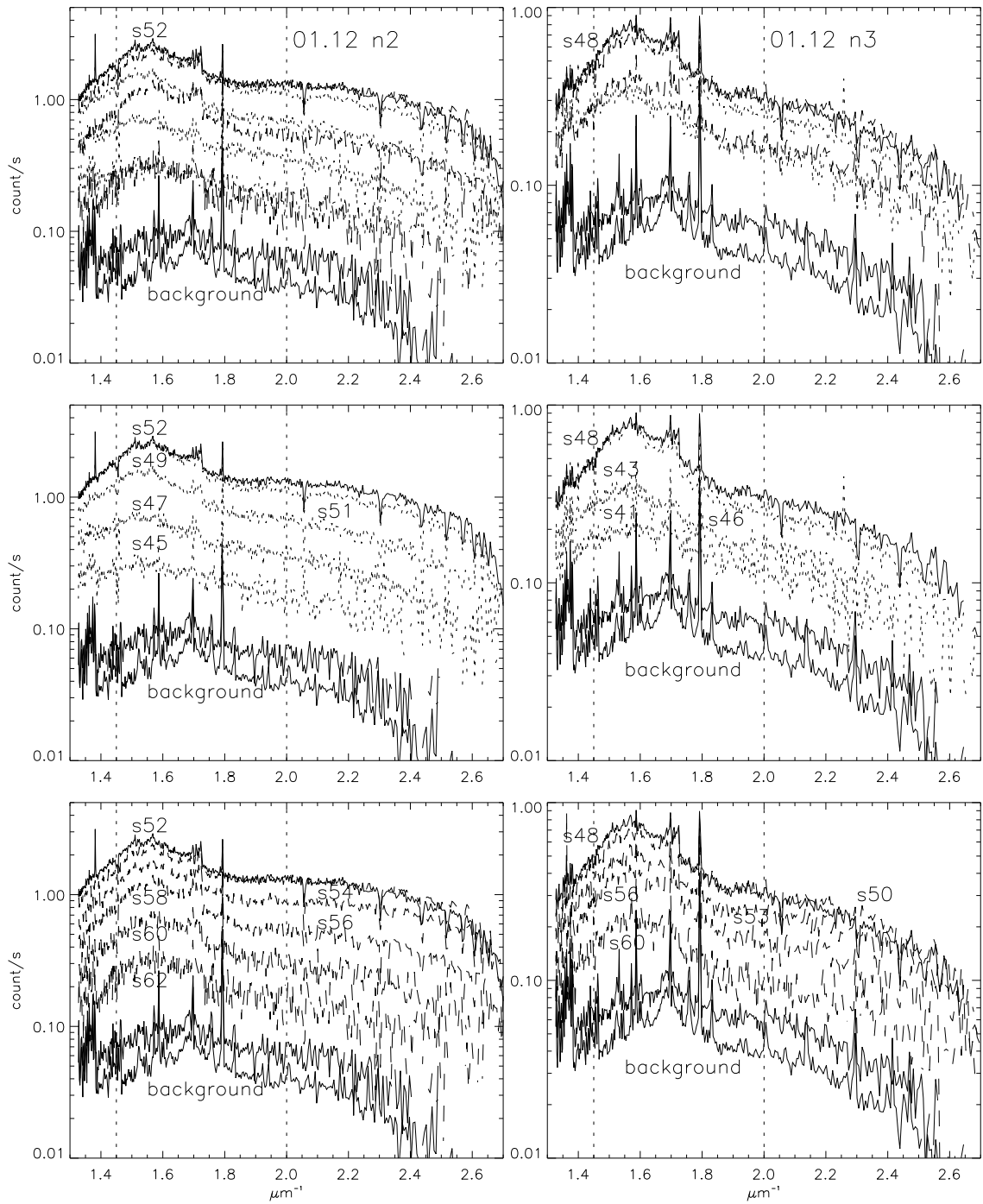}} 
\caption{December 2001 observations of the nebula at positions $n2$ (left) 
and $n3$ (right).} 
\label{fig:dec01n2n3}
\end{figure*}
\begin{figure*}[]
\resizebox{1.\textwidth}{!}{\includegraphics{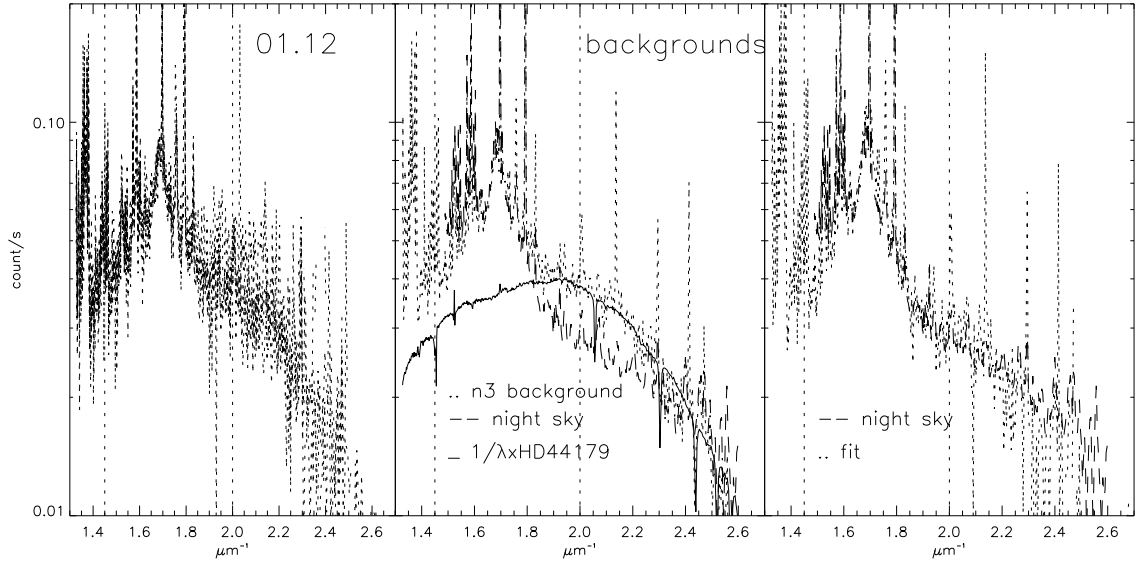}} 
\caption{\emph{Left:} backgrounds, taken at the extremities of the slit,
from December 2001 observations of the nebula ($n1$ to $n5$), 
are all identical.
\emph{Midlle:} these backgrounds correspond
to the night sky spectrum (dashes, from Massey \& Foltz 2000), 
superimposed on an additional component of light from HD44179 
scattered in the atmosphere (proportional to the spectrum of HD44179 
multiplied by $1/\lambda$), which mainly affects the blue part of the 
spectrum.
\emph{Right:} the Kitt Peak night sky is reproduced by the 
difference between the background of the nebular observation and 
a component of light from HD44179 scattered in the atmosphere.} 
\label{fig:dec01fd}
\end{figure*}
\begin{figure*}[]
\resizebox{1.\textwidth}{!}{\includegraphics{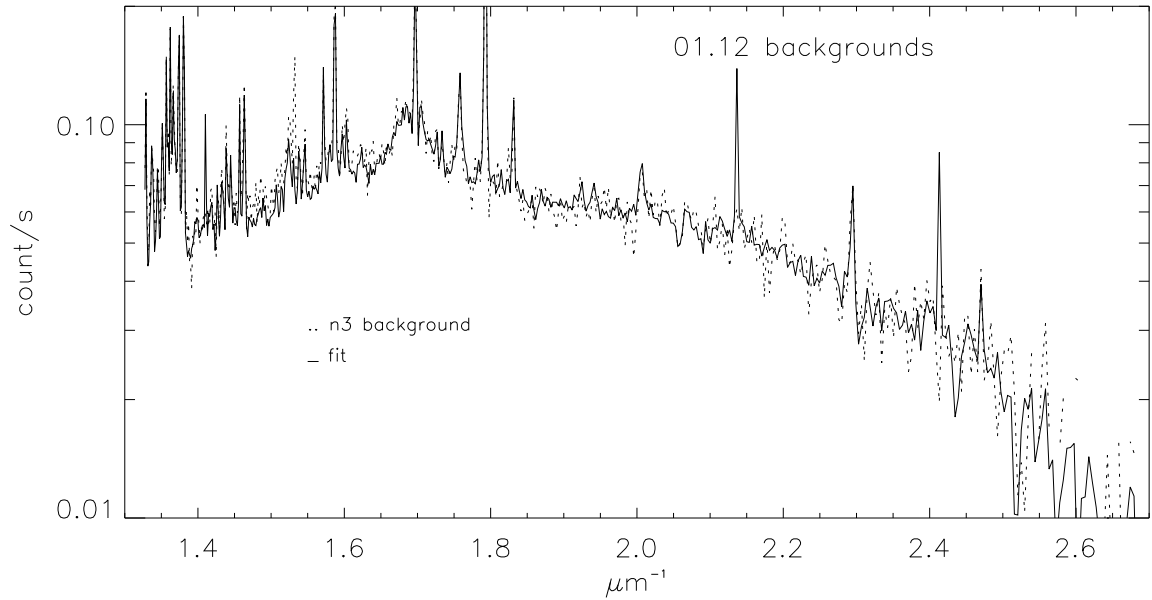}} 
\caption{The backgrounds close to the nebula are deduced from the 
backgrounds away from it by an additional component of light from HD44179 
scattered in the atmosphere. 
Here, the background, close to the nebula, 
for the December 2001 observation at position $n3$, is 
fitted by the background at the extremities of the slit, added to the 1-D spectrum of 
HD44179 multiplied by $1.2\,10^{-6}/\lambda$¥.} 
\label{fig:dec01fd1}
\end{figure*}
\begin{figure*}[]
\resizebox{1.\textwidth}{!}{\includegraphics{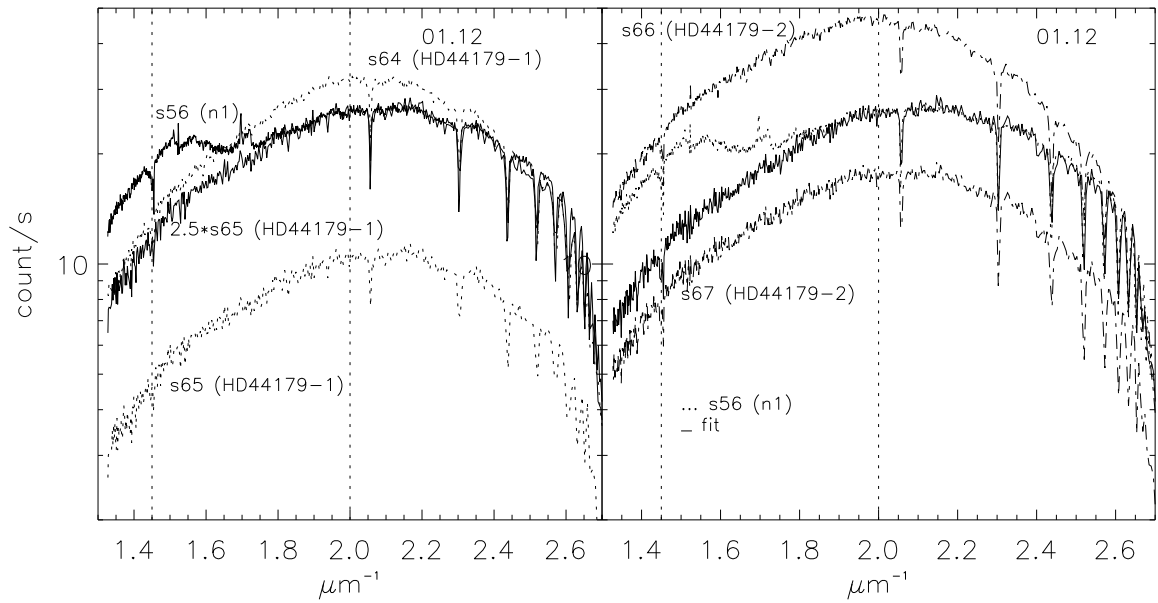}} 
\caption{\emph{Left}: main spectrum ($s56$) of $n1$ 2-D array 
is in between $s64$ and $s65$ of the first observation of HD44179, and  
is in the blue well reproduced by $2.5\times s65$.
\emph{Right}: $s56$ of $n1$ is compared to spectra of the second 
observation of HD44179. It is in-between $s66$ and $s67$, and nearly 
proportional to $s67$ (the fit is $1.3e^{0.0075/\lambda^4}s67$).
The spectrum of the nebula appears in the red only, superposed 
on the spectrum of HD44179.
} 
\label{fig:dec01n1et}
\end{figure*}
\begin{figure*}[]
\resizebox{1.\textwidth}{!}{\includegraphics{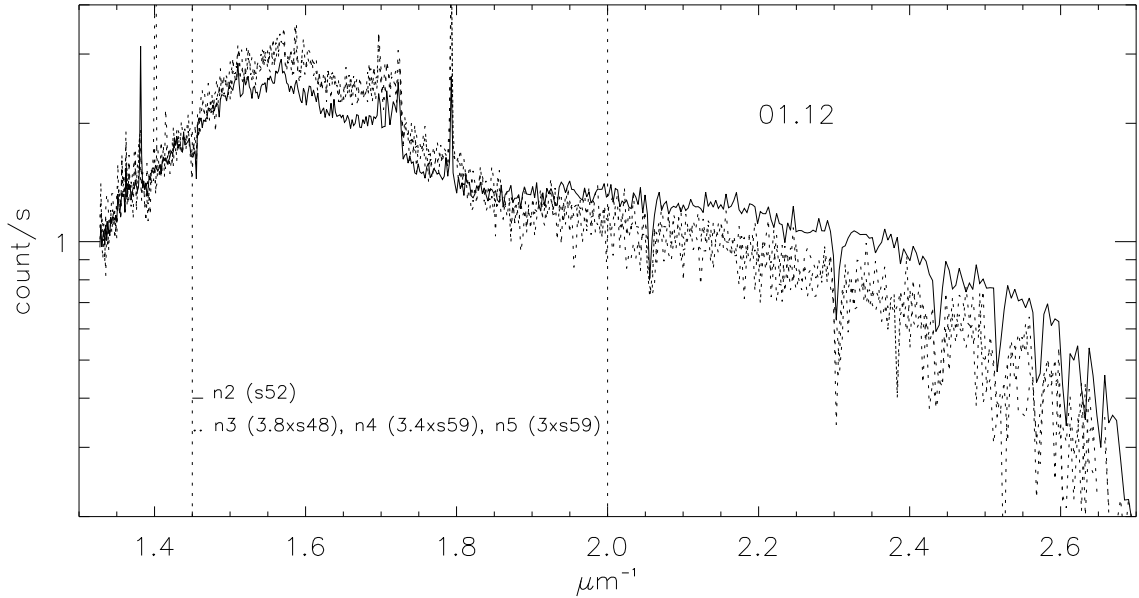}} 
\caption{Comparison of the main spectra of December 2001 observations 
of the nebula.
Positions 3, 4 and 5, in dots, differ by a constant factor only.
} 
\label{fig:dec01nc}
\end{figure*}
\begin{figure*}[]
\resizebox{1.\textwidth}{!}{\includegraphics{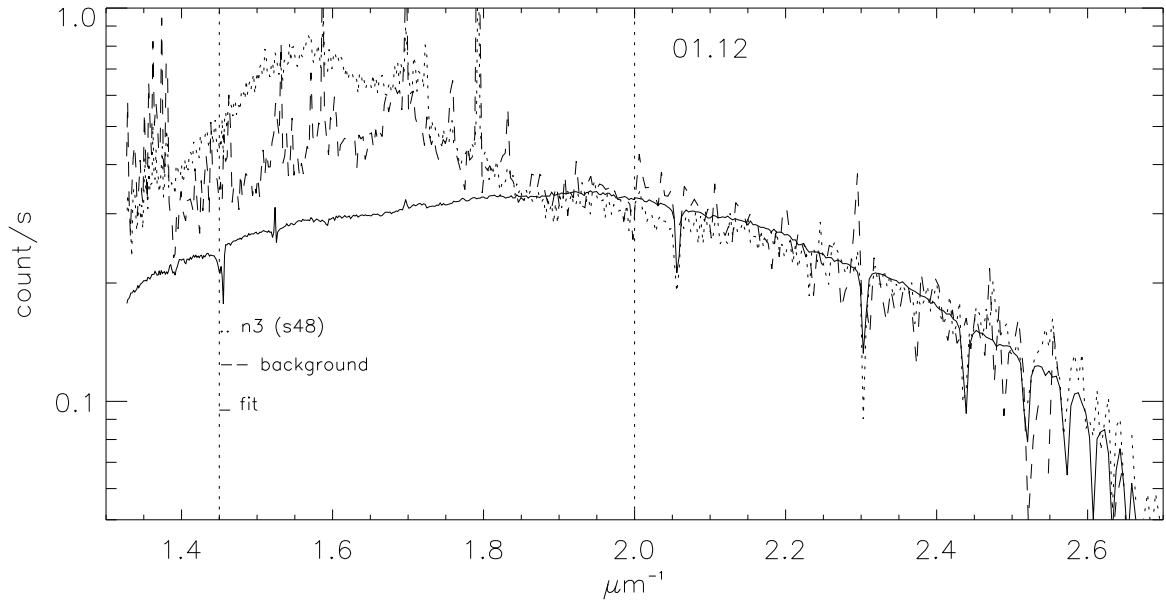}} 
\caption{Position 3 main spectrum ($s48$, dots), the background close 
to the nebula in the same observation (times 5.5, solid line), and the fit, 
proportional to the spectrum of HD44179 (first observation) and to $1/\lambda$:
the blue part of the nebular spectra $n3$, $n4$, $n5$, is light from 
HD44179 scattered in the atmosphere.} 
\label{fig:dec01n3bleu}
\end{figure*}
\begin{figure*}[]
\resizebox{1.\textwidth}{!}{\includegraphics{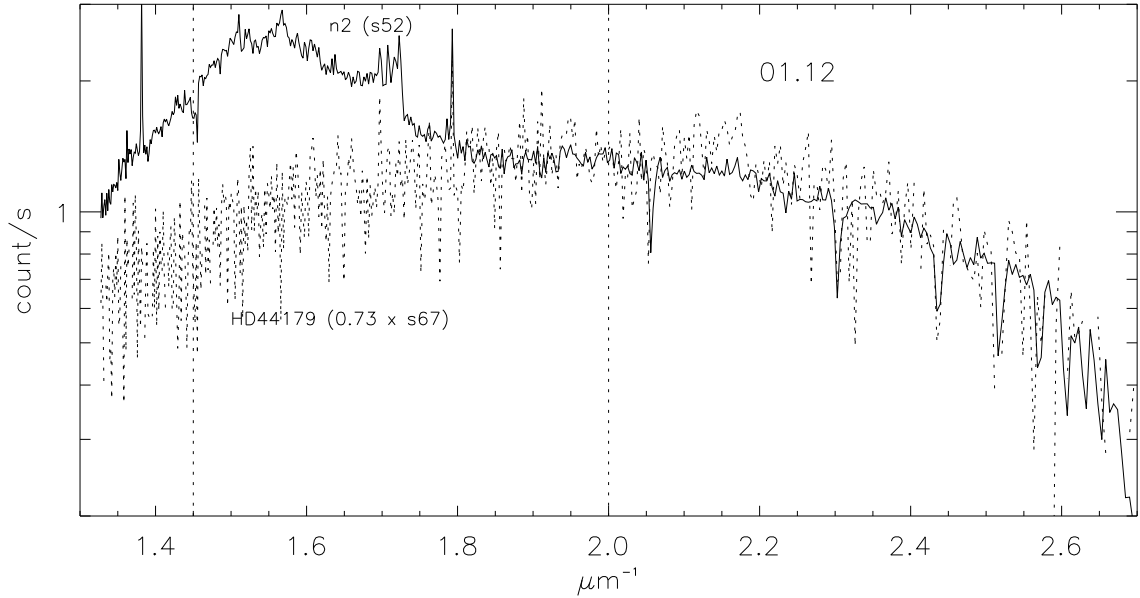}} 
\caption{The main spectrum, $s52$, from $n2$ 2-D array, is observed 
at about the same 
distance from HD44179 as $s67$ and $s68$, from the first observation 
of HD44179, are. 
The blue part of the main spectrum $s52$ at position $n2$ is 
in-between the blue parts of $s67$ and $s68$.
As shown by the plot, it is perfectly reproduced by one of these 
spectra with an appropriate scalling (here $s67$ of HD44179's first 
observation times 0.73).
} 
\label{fig:dec01n2b}
\end{figure*}
\begin{figure*}[p]
\resizebox{1.\textwidth}{!}{\includegraphics{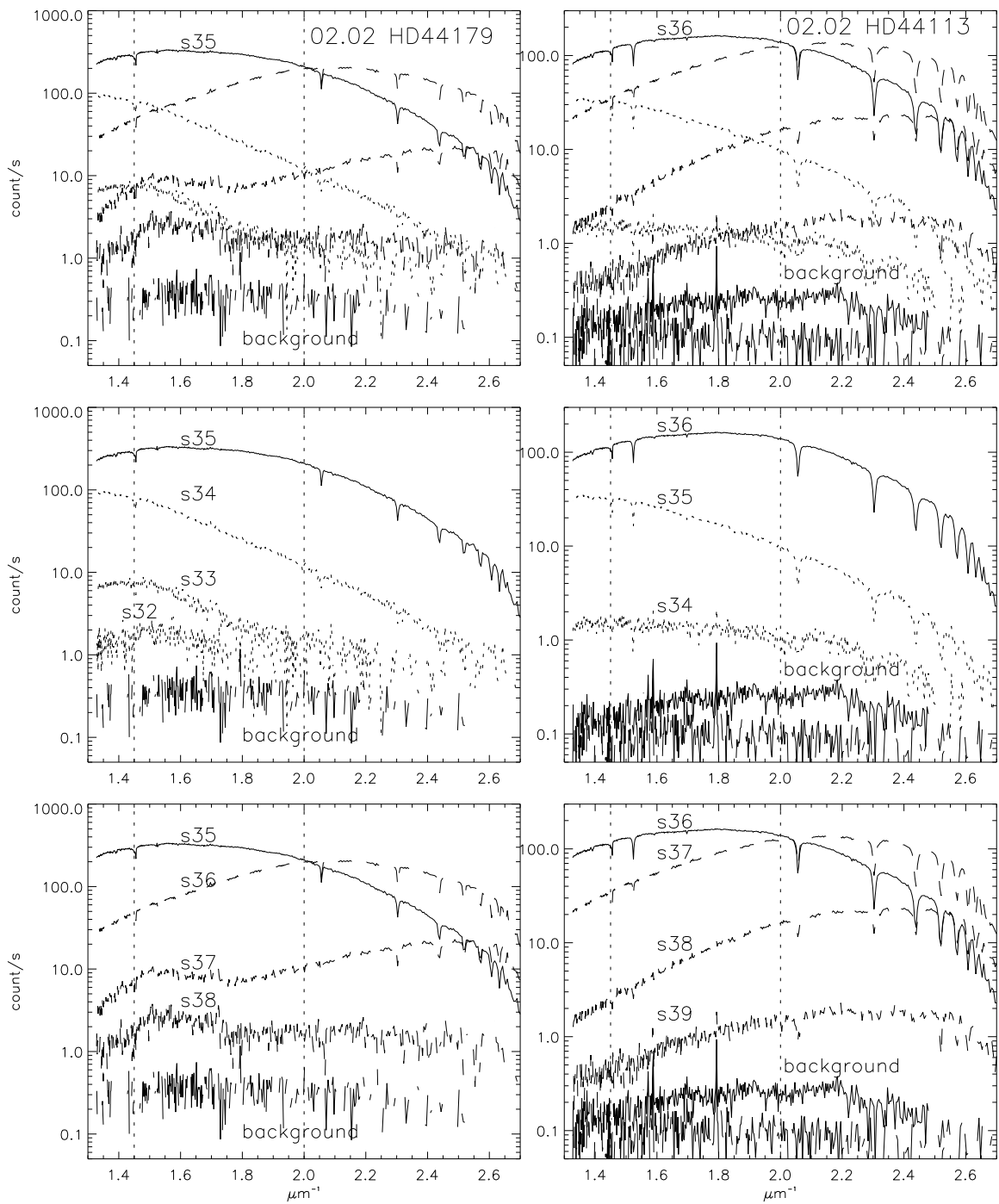}} 
\caption{February 2002 observations of HD44179 (left column) and 
HD44113 (right column).
The spectrum of the nebula appears in the red, above the spectrum of 
HD44179, for spectra $s37$, $s38$, and $s33$. Two backgrounds are 
plotted for HD44113, to illustrate its decrease with distance from the star.} 
\label{fig:feb02et}
\end{figure*}
\begin{figure*}[p]
\resizebox{1.\textwidth}{!}{\includegraphics{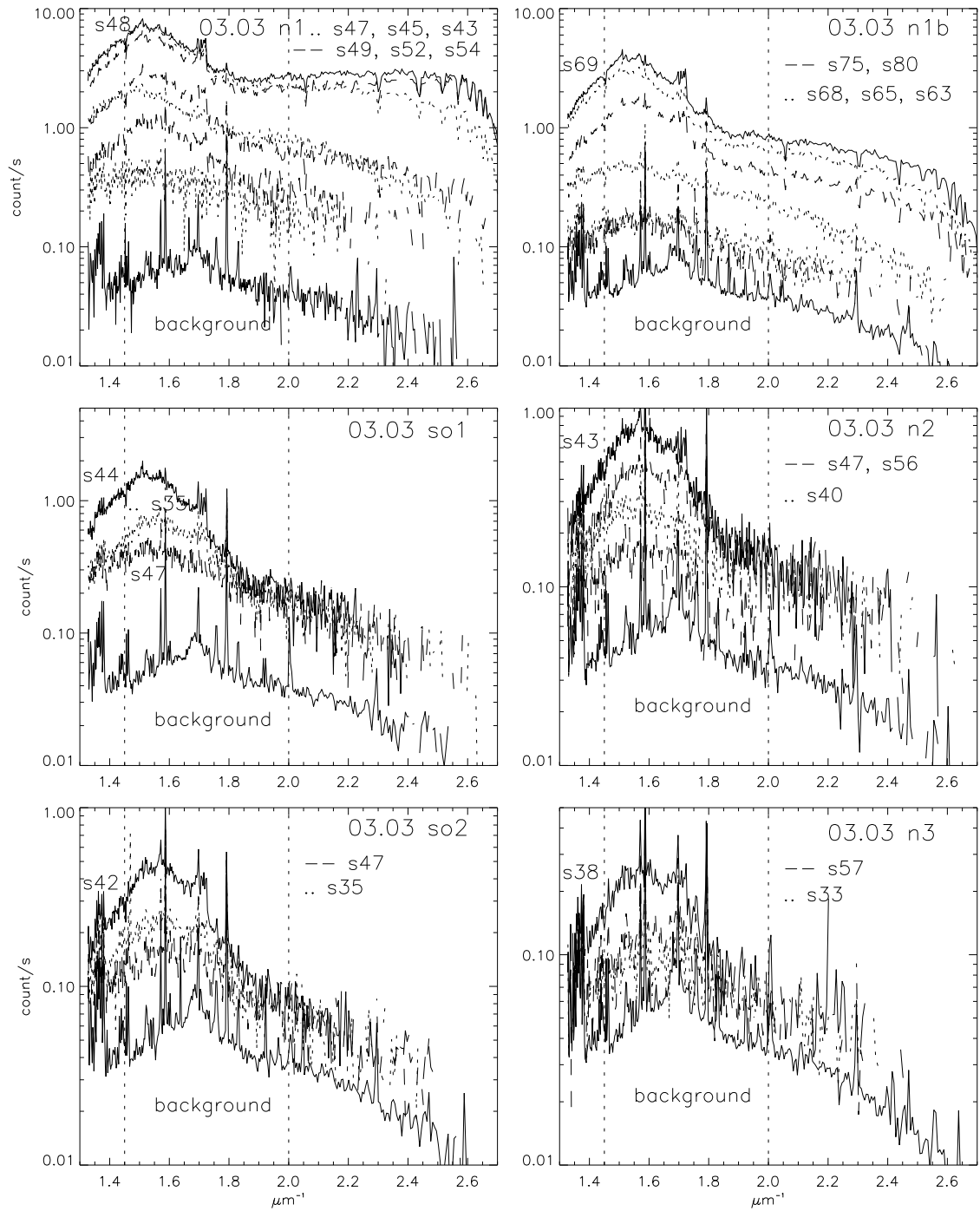}} 
\caption{March 2003 observations of the Red Rectangle nebula.
Plots are ordered from left to right and top to bottom by decreasing 
importance of the signal (i.e. by increasing distance from HD44179).
For sake of clarity, few spectra only, for each  observation, 
are represented.} 
\label{fig:ma03neb}
\end{figure*}
\begin{figure}[]
\resizebox{1.\textwidth}{!}{\includegraphics{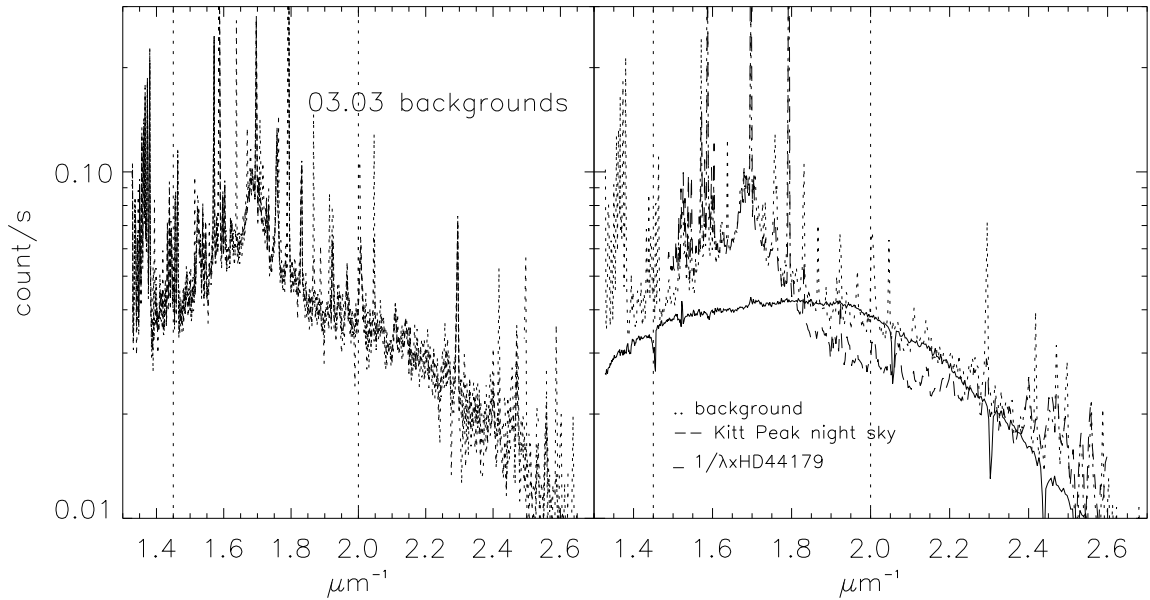}} 
\caption{\emph{Left:} The backgrounds from Figure~\ref{fig:ma03neb} (March 2003 
observations of the nebula) superimpose 
well, as for December 2001.
\emph{Right:} the night sky spectrum of Massey \& Foltz (2000) (dashes) is 
compared to the average of the background spectra (dots). 
In the blue the average spectrum presents an excess,  
due to light from HD44179 scattered in the atmosphere, fitted (solid line) by the 
spectrum of HD44179 multiplied by $1/\lambda$ (and a constant factor).} 
\label{fig:ma03fd}
\end{figure}
\begin{figure*}[]
\resizebox{\textwidth}{!}{\includegraphics{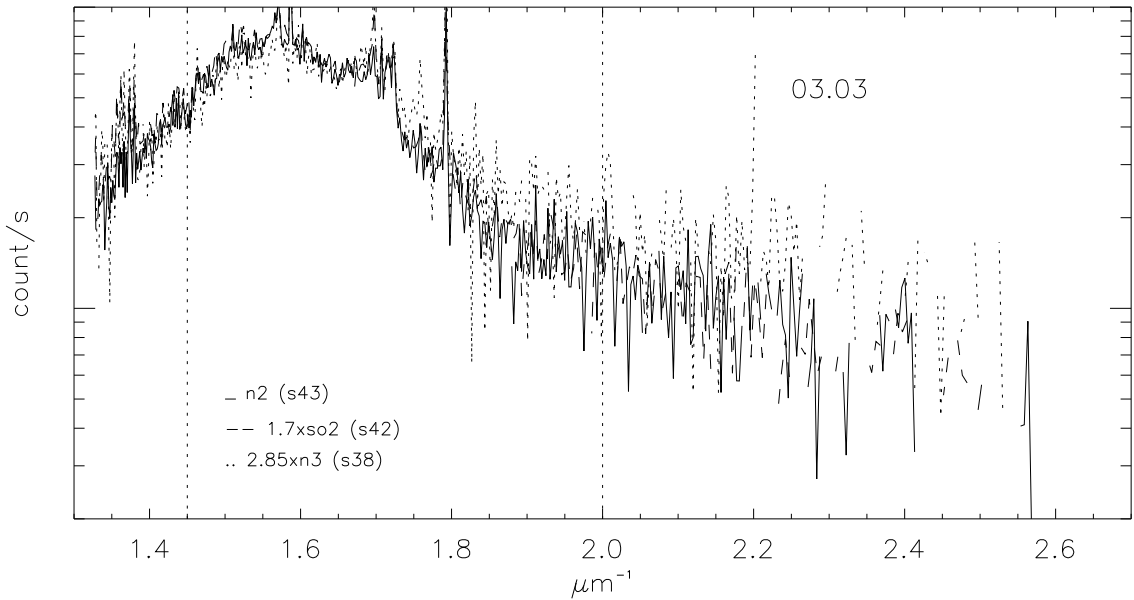}} 
\caption{The spectra at positions $n2$, $n3$, $so2$ of March 2003 
observations are identical, 
up to a constant factor ($n2=2.85\,n3=1.7\,so2$). 
}
\label{fig:ma03n2n3}
\end{figure*}
\begin{figure*}[]
\resizebox{1.\textwidth}{!}{\includegraphics{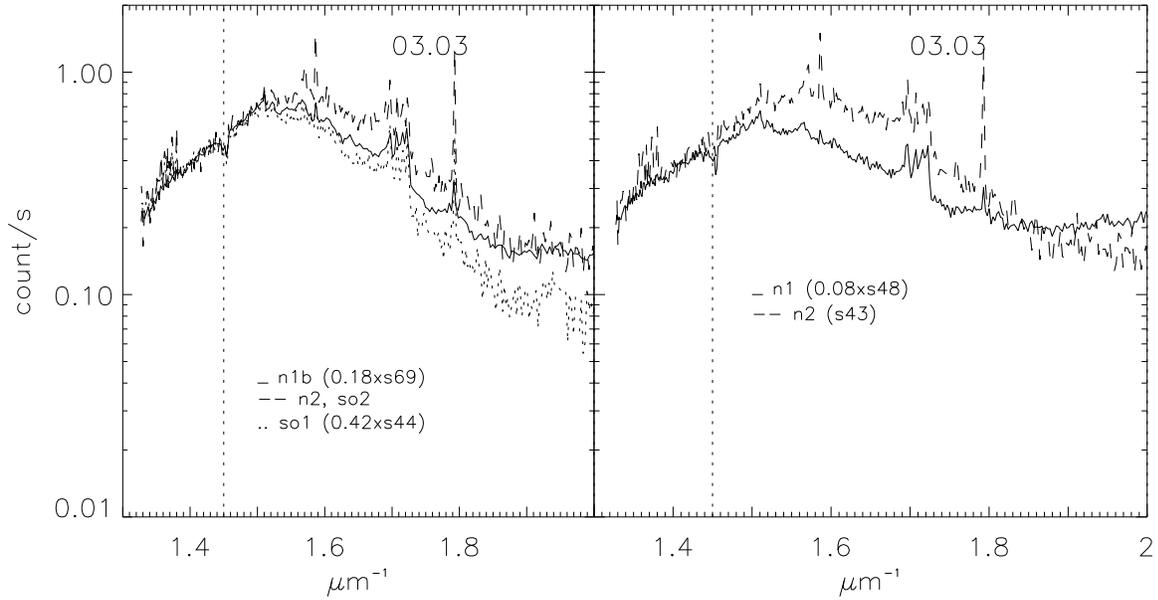}} 
\caption{\emph{Left:} The red slopes of all nebular observations 
(except $n1$) are identical. 
Here, the spectrum of the nebula averaged over positions $n2$ 
and $so2$, dashes, is compared, in the red, to the main spectra at positions 
$n1b$ (solid line) and $so1$ (dots).
\emph{Right:} the red rise of the main spectrum $s48$ at $n1$ is 
identical to the precedings in its reddest part only.
From the ozone absorption region onwards it is less steep.} 
\label{fig:ma03r}
\end{figure*}
\begin{figure*}[]
\resizebox{\textwidth}{!}{\includegraphics{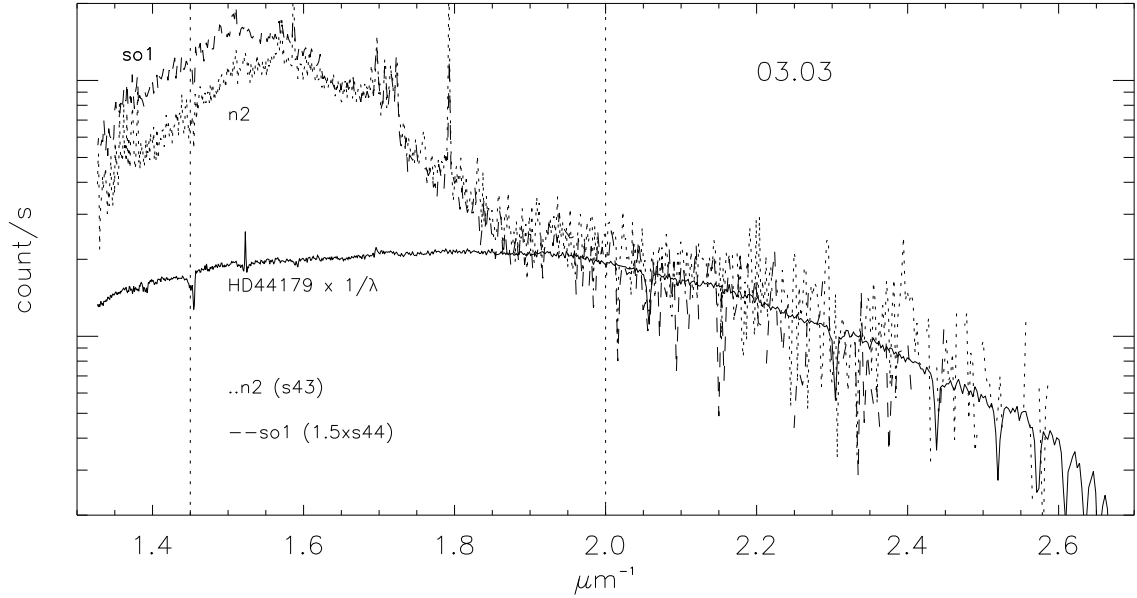}} 
\caption{Main spectra at positions $n2$ ($s43$) and $so1$ ($s44$)  are proportional in the red 
(Figure~\ref{fig:ma03r}), but also in the blue.
Their blue part reproduces the background of HD44179 observations;
it is here fitted by the spectrum of light from HD44179 
scattered in the atmosphere
(proportional to $1/\lambda$ and to the spectrum of HD44179).
} 
\label{fig:ma03n2o2}
\end{figure*}
\begin{figure}[]
\resizebox{1.\textwidth}{!}{\includegraphics{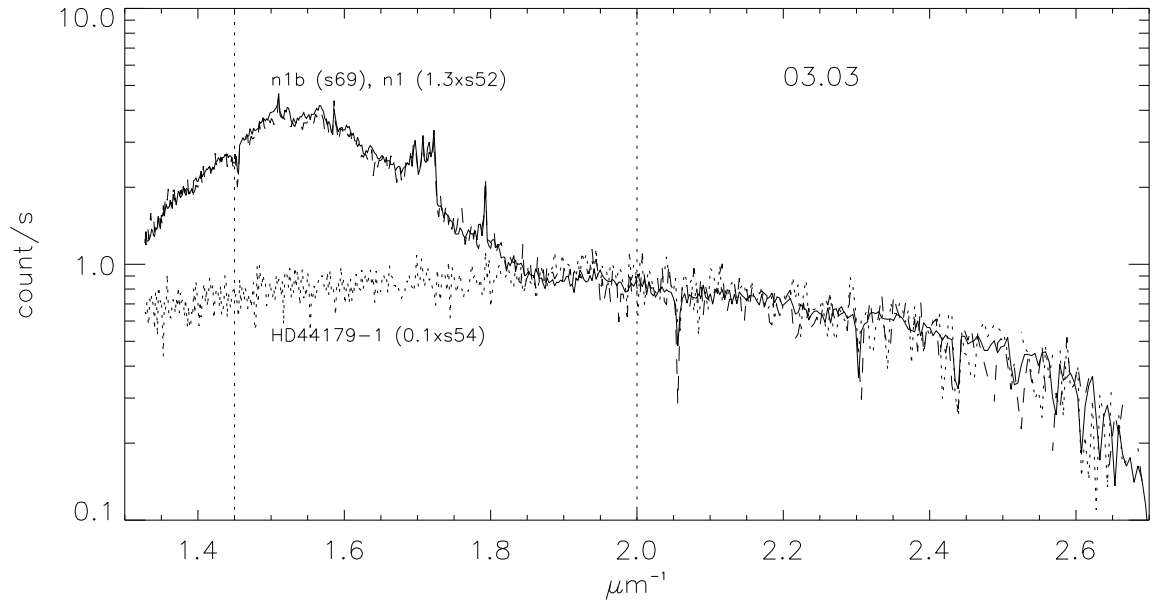}} 
\caption{The blue part of March 2003 observation $n1b$ ($s69$) is 
reproduced by spectrum $s54$ of 
the observation of HD44179 (with a decrease of 10 in intensity): 
close to the star, the spectrum of the nebula appears over the spectrum of 
refracted-scattered (in the atmosphere) light from HD44179.
The same conclusion was reached from the February observations of HD44179 
(see Figure~\ref{fig:feb02et}).
Spectrum $s52$ (dashes, scaled by a factor 1.3), from $n1$ observation, 
superimposes well on $n1b$'s $s69$.
}
\label{fig:ma03o3}
\end{figure}
\begin{figure*}[]
\resizebox{1.\textwidth}{!}{\includegraphics{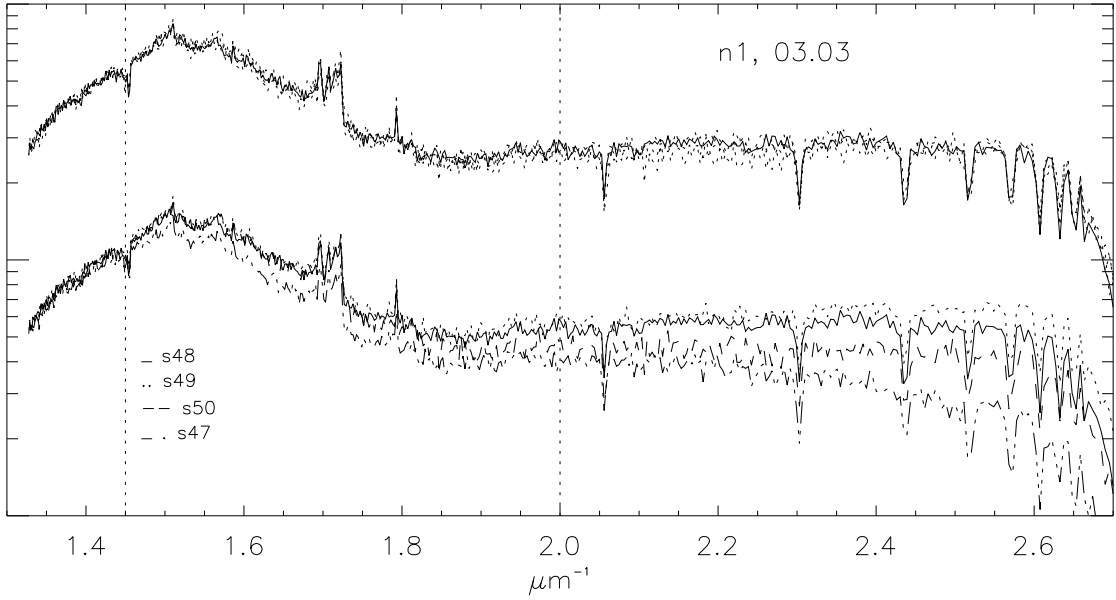}} 
\caption{\emph{Bottom spectra}: 
$s47$, $s48$, $s49$, $s50$ from $n1$ 
March 2003 observation, scaled to the same value in the red.
\emph{Top spectra}: The blue part of the spectra differ by their blue 
slope only, which is determined by the Rayleigh extinction in the 
atmosphere.
$s48$ is well reproduced by 
$0.9e^{0.018/\lambda^{4}}s47$,
$1.3 e^{-0.004/\lambda^{4}}s49$, 
$1.6e^{0.006/\lambda^{4}}s50$.
} 
\label{fig:ma03n1}
\end{figure*}
\clearpage
\begin{figure*}[]
\resizebox{\textwidth}{!}{\includegraphics{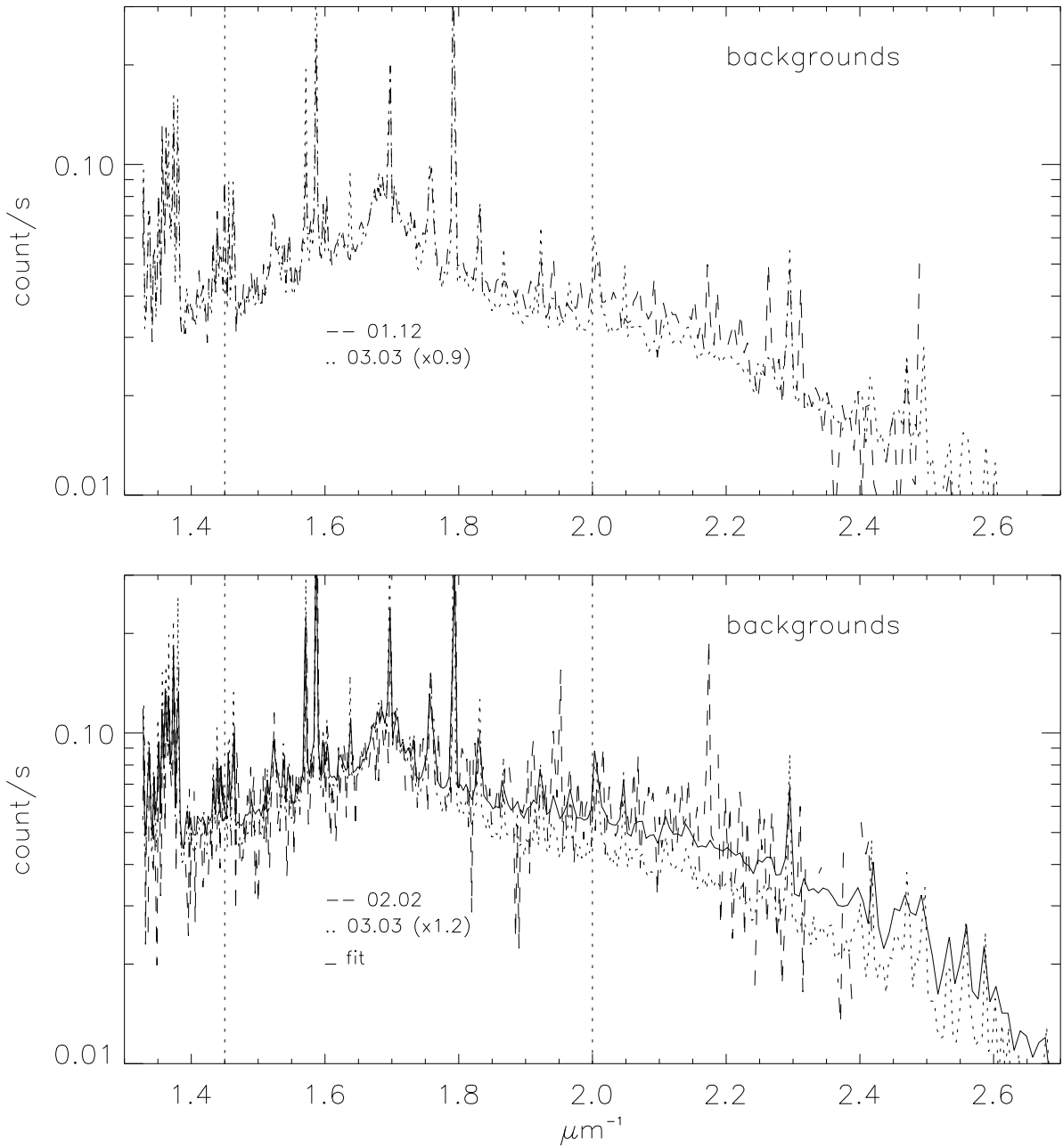}} 
\caption{Comparison of the backgrounds.
\emph{Upper plot:} December 2001 and March 2003 backgrounds are 
identical, except for a small difference in the blue.
\emph{Lower plot:} February 2002 and March 2003 backgrounds differ in the 
blue, because of the importance of the scattered light (from HD44179) 
in the atmosphere in the February background. February 2002 
background is well fitted by the sum of March 2003 background, and a 
component of light from HD44179 scattered by the atmosphere
(proportional to the February 1-D spectrum of 
HD44179 and to $1/\lambda$).
}   
\label{fig:fdc}
\end{figure*}
\begin{figure*}[]
\resizebox{\textwidth}{!}{\includegraphics{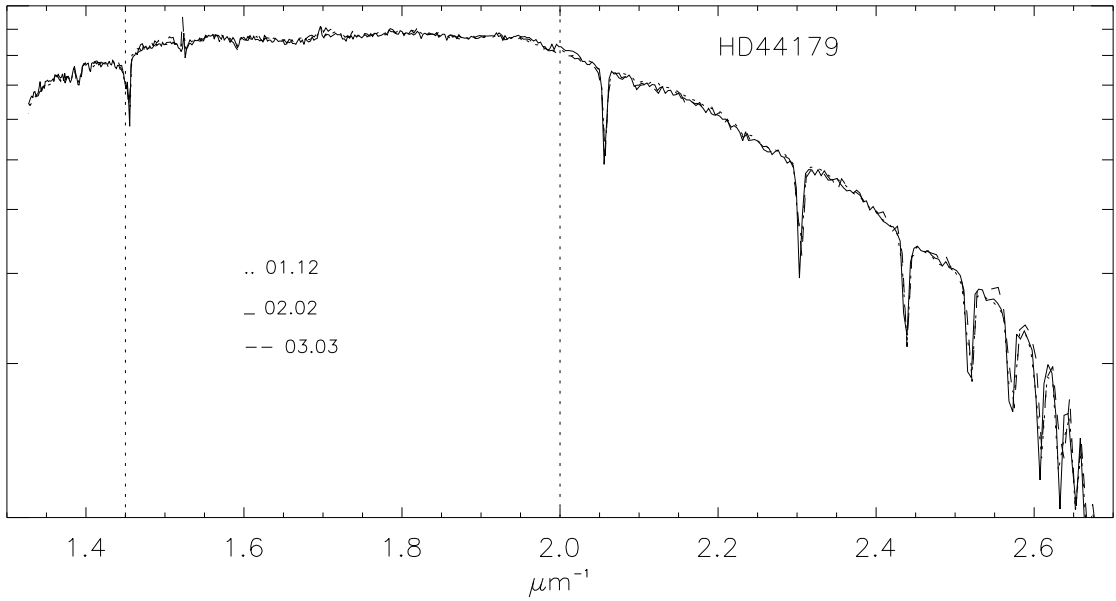}} 
\caption{The 1-D spectra of HD44179 for the three runs differ 
by an atmospheric transformation: 1-D spectra of the first observations of 
December 2001 and March 2003 superimpose on the February 2002 spectrum after 
multiplication by $1.24e^{0.008/\lambda^{4}¥}$ (for December 01), and 
$1.5e^{0.02/\lambda^{4}¥}$ (for March 03).
} 
\label{fig:etc}
\end{figure*}
\begin{figure*}[]
\resizebox{\textwidth}{!}{\includegraphics{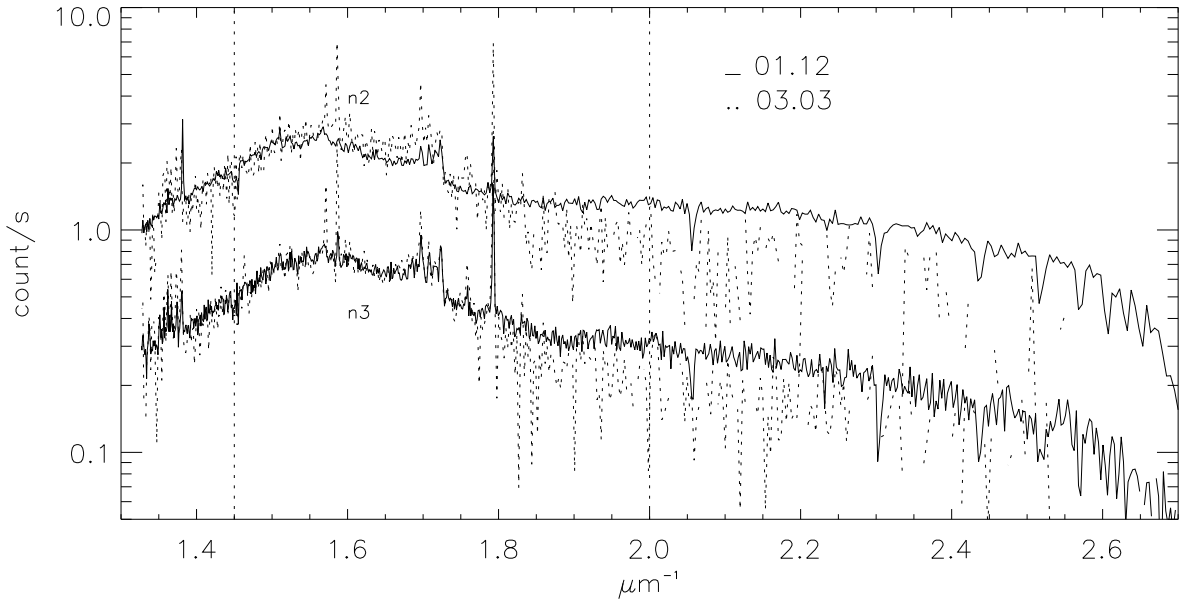}} 
\caption{Spectra observed at $n2$ and $n3$ on different dates 
are proportional in the red but differ by their blue continuum.
$n2$ and $n3$ spectra of March 2003 have been multiplied by 7 and 3 to match 
December 2001 observations.
} 
\label{fig:nloin}
\end{figure*}
\begin{figure*}[]
\resizebox{\textwidth}{!}{\includegraphics{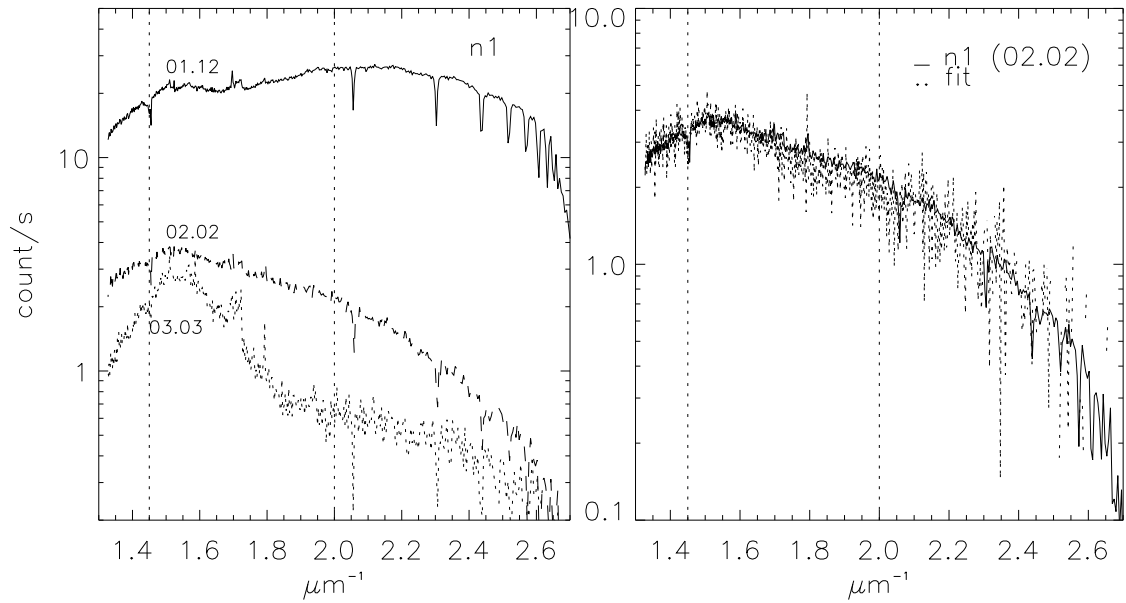}} 
\caption{Main spectra at $n1$ are extremely different from one 
date to the other (left).
The difference is explained by the position of the slit and the 
airmass, which determine the proportion of refracted and scattered 
light (in the atmosphere) from HD44179 in the 
spectrum.
The right plot shows that the main spectrum of February 2001 $n1$ 
observation can be reproduced by a combination of 
spectra from HD44179, and from the nebula at position $n2$, from 
December 2002 observation.
} 
\label{fig:n1}
\end{figure*}
\begin{figure*}[]
\resizebox{\textwidth}{!}{\includegraphics{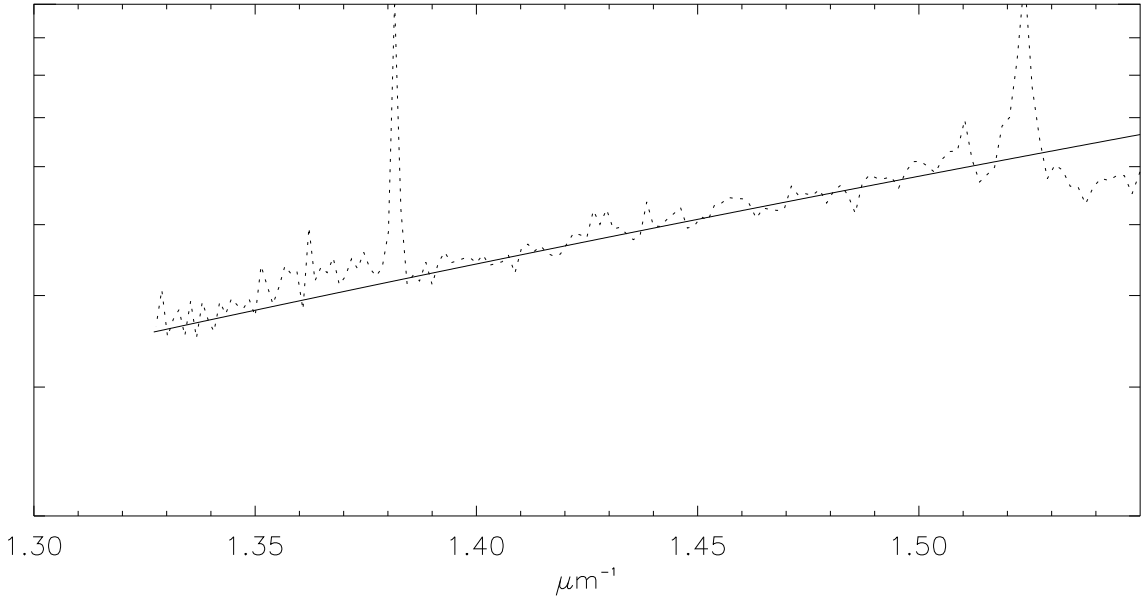}} 
\caption{In dots, the red slope of the Red Rectangle nebula
normalized by the spectrum of the A0 star HD44113 (the y-axis is 
logarithmic, the scaling is arbitrary).
It behaves as $1/\lambda^{4}$ (solid 
line), indicating scattering by the gas.
} 
\label{fig:pente}
\end{figure*}
\end{document}